\documentclass[sn-nature]{sn-jnl}% Style for submissions to Nature Portfolio journals
\usepackage{graphicx}%
\usepackage{multirow}%
\usepackage{amsmath,amssymb,amsfonts}%
\usepackage{amsthm}%
\usepackage{mathrsfs}%
\usepackage[title]{appendix}%
\usepackage{xcolor}%
\usepackage{textcomp}%
\usepackage{manyfoot}%
\usepackage{booktabs}%
\usepackage{algorithm}%
\usepackage{algorithmicx}%
\usepackage{algpseudocode}%
\usepackage{listings}%
\usepackage{comment}%
\usepackage{chemformula}%
\usepackage{multirow}
\usepackage{mhchem}
\usepackage{makecell}
\usepackage[normalem]{ulem}
\theoremstyle{thmstyleone}%
\theoremstyle{thmstyletwo}%

\theoremstyle{thmstylethree}%

\newtheorem{dfn}{Definition}[section]
\newtheorem{exa}[dfn]{Example}
\newtheorem{pro}[dfn]{Problem}

\counterwithin{dfn}{section}
\counterwithin{algorithm}{section}
\newcommand{\R}{\mathbb R}
\newcommand{\Z}{\mathbb Z}
\newcommand{\al}{\alpha}

\newcommand{\de}{\delta}

\newcommand{\la}{\lambda}
\newcommand{\La}{\Lambda}

\newcommand{\ep}{\varepsilon}

\newcommand{\angstrom}{\textup{\AA}}
\newcommand{\vol}{\mathrm{vol}}

\newcommand{\RMS}{\mathrm{RMS}}

\newcommand{\PDD}{\mathrm{PDD}}
\newcommand{\AMD}{\mathrm{AMD}}
\newcommand{\ADA}{\mathrm{ADA}}
\newcommand{\AND}{\mathrm{AND}}
\newcommand{\PDA}{\mathrm{PDA}}
\newcommand{\PPC}{\mathrm{PPC}}
\newcommand{\EMD}{\mathrm{EMD}}
\newcommand{\IS}{\mathrm{isoset}}
\newcommand{\isoset}{\mathrm{isoset}}

\newcommand{\CRDS}{\mathrm{CRDS}} % under dilation
\newcommand{\CIMS}{\mathrm{CIMS}} %under isometry 
\newcommand{\CRIS}{\mathrm{CRIS}} %under rigid motion

\newcommand{\PDDh}{\mathrm{PDD}^{\{h\}}}

\newcommand{\PDAh}{\mathrm{PDA}^{\{h\}}}
\newcommand{\PDAsq}{\mathrm{PDA}^{\{2\}}}

\newcommand{\dmin}{d_{\min}}
\newcommand{\dmax}{d_{\max}}

\newcommand{\LND}{\mathrm{LND}}

\newcolumntype{R}[1]{>{\raggedleft\let\newline\\\arraybackslash\hspace{0pt}}m{#1}}
\newcolumntype{L}[1]{>{\raggedright\let\newline\\\arraybackslash\hspace{0pt}}m{#1}}
\newcolumntype{C}[1]{>{\centering\let\newline\\\arraybackslash\hspace{0pt}}m{#1}}

\begin{document}

\title[]{Necessary conditions for benchmarking next-generation AI in materials discovery} 

%%=============================================================%%
%% Prefix	-> \pfx{Dr}
%% GivenName	-> \fnm{Joergen W.}
%% Particle	-> \spfx{van der} -> surname prefix
%% FamilyName	-> \sur{Ploeg}
%% Suffix	-> \sfx{IV}
%% NatureName	-> \tanm{Poet Laureate} -> Title after name
%% Degrees	-> \dgr{MSc, PhD}
%% \author*[1,2]{\pfx{Dr} \fnm{Joergen W.} \spfx{van der} \sur{Ploeg} \sfx{IV} \tanm{Poet Laureate} 
%%                 \dgr{MSc, PhD}}\email{iauthor@gmail.com}
%%=============================================================%%

\author[1]{\fnm{Daniel E} \sur{Widdowson}}\email{}

\author[2]{\fnm{Tatiana V} \sur{Kurlina}}\email{}
%\equalcont{These authors contributed equally to this work.}

\author[1]{\fnm{Olga D} \sur{Anosova}}\email{}
%\equalcont{These authors contributed equally to this work.}

\author[1,3]{\fnm{Vitaliy A} \sur{Kurlin}}\email{vitaliy.kurlin@liverpool.ac.uk}

%\affil*[1]{\orgdiv{Department}, \orgname{Organization}, \orgaddress{\street{Street}, \city{City}, \postcode{100190}, \state{State}, \country{Country}}}

\affil[1]{\orgdiv{School of Computer Science and Informatics}, \orgname{University of Liverpool}, \orgaddress{\street{Ashton Street}, \city{Liverpool}, \postcode{L69 3DR}, \state{Merseyside}, \country{UK}}}

\affil[2]{\orgdiv{Department of Mathematics}, \orgname{University College London}, \orgaddress{\street{Gower Street}, \city{London}, \postcode{WC1E  6BT}, \country{UK}}}

\affil*[3]{\orgdiv{Materials Innovation Factory}, \orgname{University of Liverpool}, \orgaddress{\street{51 Oxford Street}, \city{Liverpool}, \postcode{ L69 7ZD}, \state{Merseyside}, \country{UK}}}

%%==================================%%
%% sample for unstructured abstract %%
%%==================================%%

\abstract{
Current AI predicts materials by speeding up energy minimisation and training on large databases.
However, scaling data generation prevents manual analysis by domain experts. 
If novelty is evaluated by chemical compositions or traditional representations in Crystallographic Information Files, or standard structure matchers based on manually chosen thresholds, one can easily generate millions of ``new'' compositions or files without discovering truly novel materials.
\medskip

This paper uses a hierarchy of crystal invariants to continuously trace and quantify structural novelty based on the new mathematical definition of a crystal structure and asymptotic guarantees for invariant descriptors.
The invariant-based analysis through cascade computations on a modest computer exposes the unexpected scale of duplication in the world's largest materials databases, shown here in the high-profile case of GNoME datasets.
To prevent current ``embarrassment of over-prediction'', we propose necessary conditions for next-generation AI that should be automatically enforced before claiming any new materials.
}

%%================================%%
%% Sample for structured abstract %%
%%================================%%

% \abstract{\textbf{Purpose:} The abstract serves both as a general introduction to the topic and as a brief, non-technical summary of the main results and their implications. The abstract must not include subheadings (unless expressly permitted in the journal's Instructions to Authors), equations or citations. As a guide the abstract should not exceed 200 words. Most journals do not set a hard limit however authors are advised to check the author instructions for the journal they are submitting to.


%%\pacs[JEL Classification]{D8, H51}

%%\pacs[MSC Classification]{35A01, 65L10, 65L12, 65L20, 65L70}

\maketitle

\vspace*{-4mm}
\hfill
\parbox{90mm}{
\emph{Freedom is the freedom to say that two plus two make four. If that is granted, all else follows.} (George Orwell, ``1984'').} 
   
%1==============================         
\section{Introduction to challenges of materials discovery}
\label{sec:intro}

In 1988, John~Maddox’s editorial in Nature \cite{maddox1988crystals} acknowledged that “one of the continuing scandals in the physical sciences is that it remains in general impossible to predict the structure of even the simplest crystalline solids from a knowledge of their chemical composition”. 
This paper will discuss why Crystal Structure Prediction remained ``generally impossible'' before the recent developments in Geometric Data Science \cite{anosova2025geometric}. 
\smallskip

The first reason is that chemical composition may not guarantee a unique structure. 
For example, diamond and graphite consist of pure carbon but have vastly different properties.
The pharmaceutical industry is acutely aware of so-called ``late-appearing'' polymorphs whose crystal structure differs from an original prediction and might be useless for patients \cite{morissette2003elucidation}.
The broader problem is to predict, for a given composition, \emph{all possible synthesisable materials} that are stable under application conditions.  
\smallskip
 
In 2024, A.~Cheetham and R.~Seshadri \cite{cheetham2024artificial} suggested that new compounds should ``fulfill the trifecta of novelty, credibility, and utility''.
%Our paper focuses on the first novelty criterion in the case of periodic crystals without disorder.
L.~Schoop and R.~Palgrave \cite{leeman2024challenges} showed that materials discovery remains challenging because many synthesised disordered crystals are similar to known ordered structures in existing databases.
Crystallographers debated different definitions of a crystal for dozens of years and concluded in 2021 that “a real-space definition mentioning periodicity is needed'' \cite{brock2021change}.  
\smallskip

The current definition \cite{isostructural} of crystals being the ``same'' states that ``crystals are said to be isostructural if they have the same structure, but not necessarily the same cell dimensions nor the same chemical composition, and with a 'comparable' variability in the atomic coordinates to that of the cell dimensions and chemical composition''.
\smallskip

To understand what was missing in the past, recall that a \emph{periodic crystal} is traditionally described in a Crystallographic Information File (CIF) by a linear basis of three vectors in $\R^3$, usually by their lengths and pairwise angles, and a \emph{motif} of atoms with fractional coordinates in this basis.
Then a periodic crystal is obtained from a motif by periodic translations along all linear combinations of three basis vectors.
%\smallskip

\begin{figure}[h!]
\centering
\includegraphics[width=\textwidth]{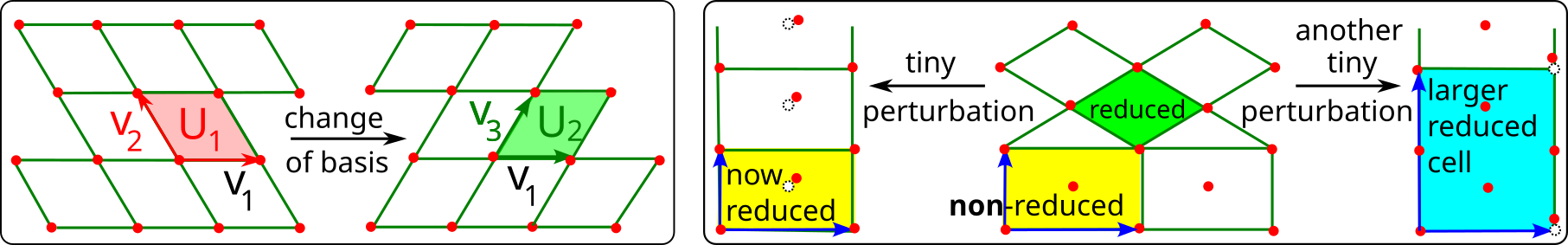}
\vspace*{0.1pt}

\includegraphics[width=\textwidth]{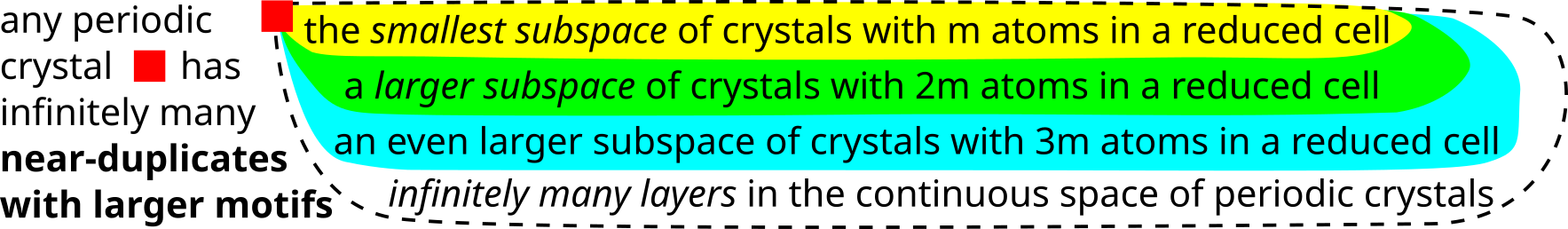}
\vspace*{-2mm}
\caption{\textbf{Top left}: a hexagonal lattice and any periodic crystal can be generated by many different cells.
\textbf{Top right}: a reduced cell can discontinuously change under tiny trivial perturbations.
\textbf{Bottom}: any periodic crystal belongs to infinitely many subspaces of crystals with motifs of multiple sizes.}
\label{fig:basis_ambiguity}
\end{figure}

Any basis generates the \emph{unit cell}, but any periodic crystal can be represented by infinitely many pairs (unit cell, motif).
Though a unit cell can be reduced to a canonical form, e.g. Niggli's cell \cite{niggli1928krystallographische}, it has been experimentally known since 1965 that Niggli's cell discontinuously changes under noise  \cite[p.~80]{lawton1965reduced}, see
%An example in 
Fig.~\ref{fig:basis_ambiguity}~(top right). % extends to any crystal.
%Indeed, already in dimension 1 the integer sequence $\Z$ of period 1, for any small $\ep>0$, is infinitesimally close to the sequence $\Z(\ep;m)=\{\ep,1,\dots,m\}+(m+1)\Z$ with an arbitrarily larger period $m+1$.  
\smallskip

The ``same structure'' remained undefined in \cite{isostructural} until a \emph{crystal structure} was introduced in \cite[Definition~6]{anosova2024importance} as a class of all crystals that can be exactly superimposed with all atoms by \emph{rigid motion}, which is a composition of translations and rotations. % in $\R^3$.
\smallskip

This definition does not allow any `comparable' variability of crystal parameters.
Indeed, if we allow all perturbations up to any small threshold $\ep>0$, sufficiently many trivial perturbations can make all crystals equivalent due to the transitivity axiom: if a crystal $S$ is equivalent to a crystal $Q$, and $Q$ is equivalent to a crystal $T$, then $S$ is equivalent to $T$, briefly written as follows: if $S\sim Q\sim T$ then $S\sim T$. 
\smallskip

The ambiguity under noise, illustrated in Fig.~\ref{fig:basis_ambiguity}~(top right), has been known since ancient times as the \emph{sorites paradox}: ``When does a heap of sand stop being a heap while grains of sand are removed one by one?'' \cite{oms2019introduction}. 
Since atoms have real coordinates and are not restricted to a discrete grid, atomic positions are like ``grains of sand'' that can be slightly shifted again and again until a crystal becomes completely different.
Using tolerances for a binary outcome (true/false) of matching %crystals or other real objects 
only shifts the comparison from 0 to another threshold $\ep>0$ without resolving the sorites paradox.
\smallskip

For continuous classifications of real data, we can distinguish any tiny differences by a continuous distance metric $d$ so that if periodic crystals $S,Q$ are related by a small perturbation moving each atom up to a Euclidean distance $\ep$, then their distance $d(S,Q)$ is small relative to $\ep$.
If there is a constant $\la$ such that $d(S,Q)\leq\la\ep$ for all perturbations up to a small $\ep>0$, this condition is called \emph{Lipschitz continuity}.
%we can measure distances between atoms and then crystals continuously without splitting them into discrete classes by symmetries. 
%For example, a natural distance between the equivalence classes of the periodic sequences $\Z$ and $\Z(\ep;m)$ of very different periods $1$ and $m+1$ should be close to 0 and depend continuously on a small $\ep>0$.
\smallskip

%Past approaches to manually fix hard thresholds for similarity of continuous data  \smallskip

The continuous nature of our world necessarily requires us to distinguish between all (infinitely many) rigid structures by continuous metrics.
How is it really possible? 
\smallskip

Since crystal structures are determined in a rigid form without a preferred coordinate system, there is no sense in distinguishing between their different representations $S\neq Q$ that are \emph{rigidly equivalent}, i.e. $S$ can be exactly matched with $Q$ by rigid motion, denoted as $S\cong Q$.
This rigid equivalence is an example of an \emph{equivalence relation} (denoted as $\sim$) that should satisfy three basic axioms: (1) \emph{reflexivity}: $S\sim S$, (2) \emph{symmetry}: if $S\sim Q$ then $Q\sim S$, and (3) \emph{transitivity}: if $S\sim Q\sim T$, then $S\sim T$.
The last axiom is needed for any justified classification as a splitting into disjoint classes.
\smallskip
 
Any periodic crystal $S$ generates its \emph{crystal structure} in  \cite[Definition~6]{anosova2024importance} as the equivalence class $[S]=\{\text{all crystals } Q\cong S\}$ consisting of all (infinitely many) crystals $Q$ that are rigidly equivalent to a single representative $S$ in a CIF.
The resulting space of crystal structures is continuous because any perturbation of a crystal $S$ almost always produces a slightly different structure $[Q]\neq[S]$, as in Fig.~\ref{fig:basis_ambiguity}~(top right).
\smallskip

The rigid equivalence can be extended by adding compositions with mirror reflections (\emph{isometry}) and uniform scaling (\emph{dilation}), which makes the resulting spaces smaller by not distinguishing between mirror images, but keeps them continuous.
If we view atoms as zero-sized points without chemical types, all simple cubic crystals of different properties, such as rock salt \ce{NaCl} or \ce{KBr}, become equivalent under dilation. 
\smallskip

Hence, if we aim to distinguish between all functionally different (periodic) materials, we should study them in the largest continuous space, without splitting it into disjoint pieces by symmetries or chemical compositions.
This huge space of all theoretical periodic structures 
%based on points without elements 
opened the problem of finding a mathematical code  %\cite[Problem~10]{anosova2024importance} 
or a \emph{materials genome} \cite{de2019new} that should identify any material with all its properties \cite{nosengo2016material}.
%The next section will summarise progress towards such a code whose absence prevents justified materials prediction.

%2====================
\section{Towards a materials genome of periodic crystals}
\label{sec:definitions}

A practical approach to resolve the \emph{curse of dimensionality} in the continuous crystal space is to quickly narrow the search in a hierarchical way until a unique material is reliably identified.
While many real objects, including molecules and crystals, are still compared by aligning their geometric representations in $\R^3$, Emmy Noether \cite{rowe2021emmy} initiated a more general and efficient invariant-based approach as early as 1918 \cite{noether1918invariant}. 
\smallskip

For a fixed equivalence relation, such as rigid motion, an \emph{invariant} $I$ is a function that takes the same value on all equivalent objects.
In other words, if crystals $S\cong Q$ are rigidly equivalent, a rigid invariant should satisfy $I(S)=I(Q)$.
A simple invariant is the minimal volume $\vol(S)$ of a (primitive) unit cell of a crystal $S$. 
Instead of computationally expensive alignment, we can conclude that crystals $S,Q$ are not rigidly equivalent if any (single) invariant $I$ takes different values on them: $I(S)\neq I(Q)$.
\smallskip

If an invariant $I$ satisfies $I(S)=I(Q)$, we cannot conclude that $S\cong Q$ in general.
If this inverse property always holds (if $I(S)=I(Q)$ then $S\cong Q$), the invariant $I$ is called \emph{complete}.
The standard settings of a periodic crystal with a reduced cell form a complete invariant in theory \cite{parthe2013typix}.
In practice, any reduced cell discontinuously changes under almost any noise, as in Fig.~\ref{fig:basis_ambiguity}~(top right).
The practical requirement of continuity for periodic structures was introduced in 2020 \cite{mosca2020voronoi} and led to a hierarchy of continuous invariants from computationally fast vectors \cite{widdowson2022average} to complete ones \cite{anosova2026recognition}. 
\smallskip

For a periodic crystal $S\subset\R^3$ and any integer $k\geq 1$, 
%with points $p_1,\dots,p_m$ in a unit cell, 
our first invariant $\AMD_k(S)$, called the \emph{average minimum distance}, is the average distance to the $k$-th atomic neighbour in $S$, see Fig.~\ref{fig:AMD}.
The sequence of $\AMD_k(S)$ is computed up to a fixed number $k$ of atomic neighbours within the full crystal $S$, without restricting to any unit cell.
%\smallskip

\newcommand{\amdh}{46mm}
\begin{figure*}[h!]
\centering
\includegraphics[height=\amdh]{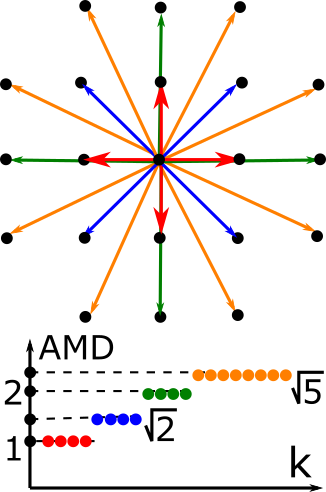}
\hspace*{3mm}
\includegraphics[height=\amdh]{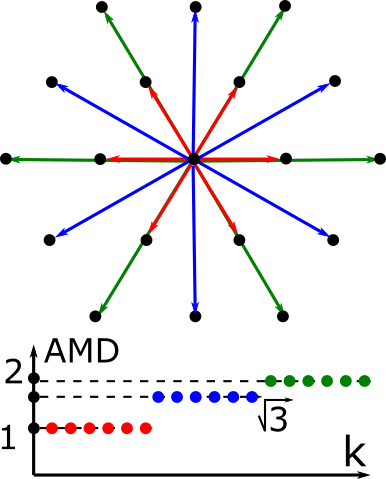}
\hspace*{3mm}
\includegraphics[height=\amdh]{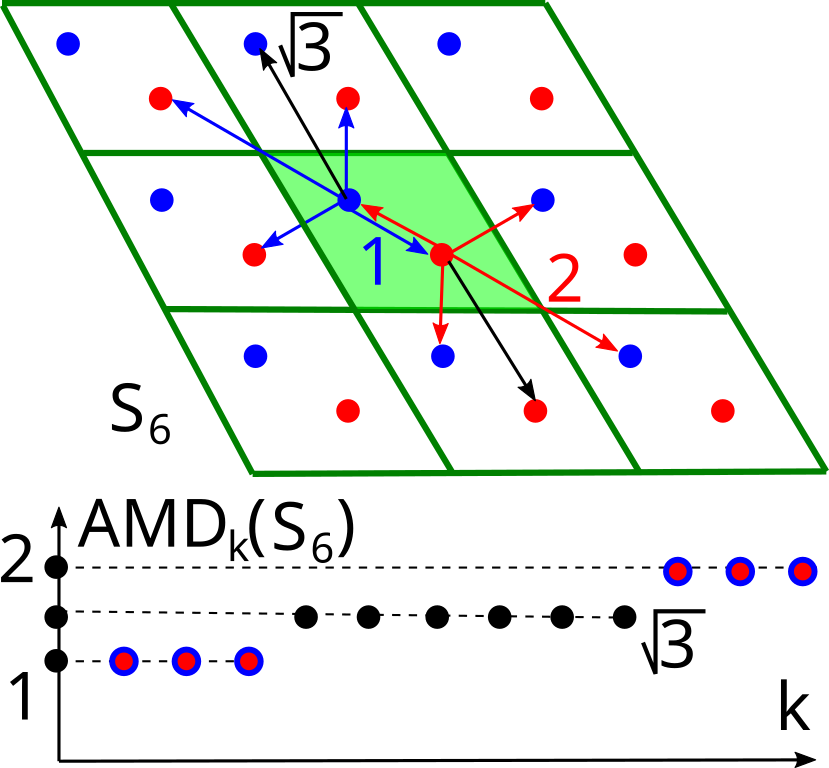}
\caption{\textbf{Left}: in the square lattice, the $k$-th neighbours of the origin and corresponding $\AMD_k$  are shown in the same colour. For example, the shortest axis-aligned distances $\AMD_1=\dots=\AMD_4=1$ are in red, the longer diagonal distances $\AMD_5=\dots=\AMD_8=\sqrt{2}$ are in blue.
\textbf{Middle}: in the hexagonal lattice, the shortest distances are in red: $\AMD_1=\dots=\AMD_6=1$.
\textbf{Right}: a honeycomb periodic point set has a 2-point motif and the first average distances $\AMD_1=\AMD_2=\AMD_3=1$.} 
\label{fig:AMD}
\end{figure*}

More traditional atomic environments up to a fixed cut-off radius $r$ can contain different numbers of atoms and change discontinuously when atoms are perturbed across a hard boundary.
While the order of equidistant neighbours can change under noise, their ordered distances to a central atom are Lipschitz continuous in the sense that perturbing any points $p,q$ up to $\ep$ changes their distance $|p-q|$ up to $2\ep$.
\smallskip

The last conclusion follows for any distance \emph{metric} $d$ that satisfies three axioms for all objects under a given equivalence: (1) $d(S,Q)=0$ if and only if $S\sim Q$, (2) \emph{symmetry}: $d(S,Q)=d(Q,S)$, and (3) \emph{triangle inequality} $d(S,Q)+d(Q,T)\geq d(S,T)$.
If the triangle inequality fails, outputs of the clustering algorithms $k$-means and DBSCAN may not be trustworthy \cite{rass2024metricizing}.
The \emph{only if} condition in the first axiom guarantees completeness because any non-equivalent crystals $S\not\sim Q$ have $d(S,Q)>0$.  
\smallskip

In all our invariants, the number $k$ of atomic neighbors is not treated as a parameter, but as a degree of resolution, similar to the number of decimal places on a calculator. 
Indeed, as $k$ increases, the sequence of $\AMD_k(S)$ expands by considering more distant atomic neighbors while all previous invariants remain unchanged.
\smallskip

The key advantage of the invariant $\AMD(S;k)=(\AMD_1(S),\dots,\AMD_k(S))$ and its stronger extensions below is Lipschitz continuity: if periodic crystals $S,Q$ are related by shifting every atom up to $\ep$, the Root Mean Square distance has the upper bound 
$$\RMS\big(\AMD(S;k),\AMD(Q;k)\big)=\sqrt{\dfrac{1}{k}\sum\limits_{i=1}^k\big(\AMD_i(S)-\AMD_i(Q)\big)^2}
\leq 2\ep \text{ for } k\geq 1.$$
The upper bound $2\ep$ explains the physical meaning of $\RMS$, which is measured in the same units as point coordinates: if $\RMS(\AMD(S;k),\AMD(Q;k))=d$, to exactly match $S,Q$ by isometry, we should perturb the first $k$ inter-atomic distances by $d$ on average.
Averaging over all the atoms in a motif produces an invariant that is independent of point ordering but theoretically incomplete.
Instead of averaging, the Pointwise Distance Distribution ($\PDD$), for each atom $p$ in a motif, keeps the first $k$ distances to its neighbours in a separate list \cite{widdowson2022resolving}, so the full invariant $\PDD(S;k)$ is an unordered distribution of these equally weighted lists, see details in Definition~\ref{dfn:PDD}.
\smallskip

For a crystal $S$ with a motif of $m$ atoms, one can initially write $\PDD(S;k)$ as a matrix with $k$ ordered columns and $m$ unordered rows.
The honeycomb structure $S_6$ in Fig.~\ref{fig:AMD}~(right) has $m=2$ points that can be matched (with the full $S_6$) by a global rotation.
As a matrix, $\PDD(S_6;k)$ for $k\geq 4$ has two identical rows starting with distances $1,1,1,\sqrt{3},\dots$.
It is convenient to collapse all identical (say, $l$) rows to a single one with the weight $\frac{l}{m}$, so $\PDD(S_6;k)$ becomes a single vector of weight 1.
\smallskip

Interpreting $\PDD(S;k)$ as a probability distribution of vectors in $\R^k$, we often apply the Earth Mover's Distance ($\EMD$) from transportation theory \cite{kantorovich1960mathematical}, which optimises the cost of transforming distance lists in two PDDs by using $\RMS$ on $\R^k$.
\smallskip

Due to the Lipschitz continuity, if $\EMD\big(\PDD(S;k),\PDD(Q;k)\big)=d$, to match the crystals $S,Q$ by isometry, atoms of $S$ should be mapped to atoms of $Q$ so that the first $k$ inter-atomic distances are perturbed by $d$ overall \cite[Theorem~4.2]{widdowson2026pointwise}. 
In other words, $S$ and $Q$ cannot be exactly matched if we shift every atom by less than $\frac{d}{2}$.
\smallskip

More significantly, $\PDD(S;k)$ with a large enough $k$ and a lattice of $S$ suffices to reconstruct almost any periodic crystal $S\subset\R^3$, uniquely under isometry, see \cite[Theorem 5.5]{widdowson2026pointwise}.
Although this \emph{generic completeness} does not guarantee completeness in theory, we have not found any non-isometric structures (denoted as $S\not\simeq Q$) with $\PDD(S;100)=\PDD(Q;100)$ among all experimental crystals in modern databases.
\smallskip

In 1930, Linus Pauling \cite{pauling1930crystal} found non-isometric crystals that had identical diffraction patterns and were later called \emph{homometric} \cite{patterson1939homometric}.
Second-order PDD invariants distinguished the Pauling crystals and all other theoretically known periodic sets $S\not\simeq Q$ with $\PDD(S;k)=\PDD(Q;k)$ for $k\geq 1$ in \cite[Examples~3.4 and 4.2]{widdowson2027higher}.
\smallskip

For a periodic crystal $S\subset\R^3$ with $m$ atoms in a unit cell of volume $\vol[U]$, the \emph{Point Packing Coefficient} $\PPC(S)=\sqrt[3]{\dfrac{\vol[U]}{mV_3}}$ measures the (scaled) volume per atom, where $V=\dfrac{4}{3}\pi$ is the unit ball volume in $\R^3$.
\cite[Theorem~13]{widdowson2022average} proved that 
the distance to the $k$-th atomic neighbour in $S$ is asymptotically increasing as $\PPC(S)\sqrt[3]{k}$ as $k\to+\infty$.
Hence, if periodic crystals $S,Q$ have $\PPC(S)\neq\PPC(Q)$, the invariants $\AMD_k(S)$ and $\AMD_k(Q)$ eventually diverge, so their $\RMS$ is dominated by contributions of the most distant neighbours.
However, the most interesting differences appear for small $k$, such as the average bond length in molecular crystals for $k=1$.
\smallskip

\begin{figure}[h!]
\centering
\includegraphics[width=\textwidth]{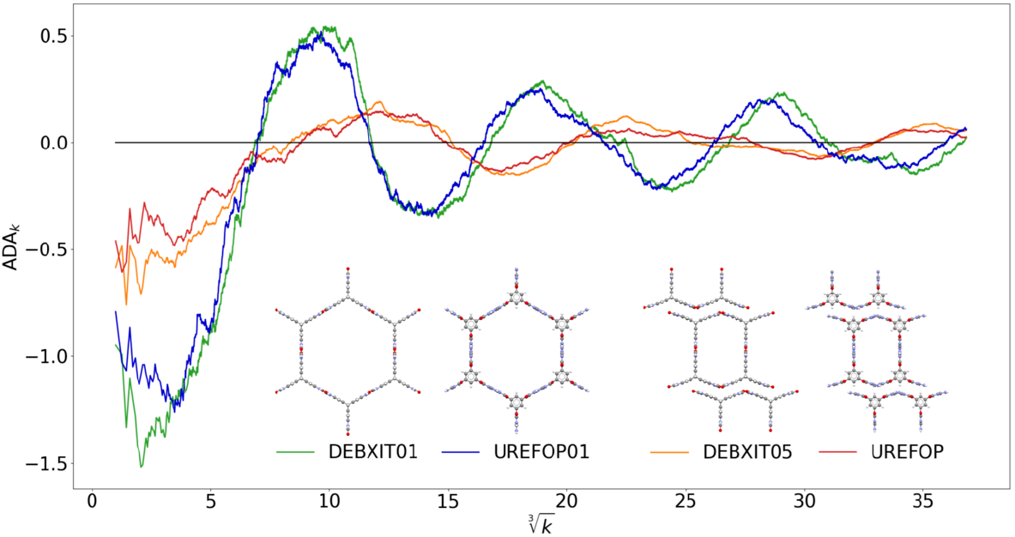}
\vspace*{-4mm}
\caption{The invariants $\ADA_k$ reveal the similarity between some crystals %(DEBXIT vs UREFOP01, and DEBXIT05 vs UREFOP) 
in the Cambridge Structural Database (CSD), though the molecules in crystals DEBXIT and UREFOP differ in size and chemistry. }
\label{fig:ADA-waves}
\end{figure}

To minimise this effect, we adjusted $\AMD_j(S)$ and all distances in the $j$-th column of $\PDD(S;k)$ for $j\leq k$ by subtracting $\PPC(S)\sqrt[3]{j}$.
The resulting invariants $\ADA(S;k)$ and $\PDA(S;k)$ are called \emph{average} and \emph{pointwise deviations from asymptotic}, respectively.
For molecular crystals, $\ADA_k$ behaves like waves fading to $0$ in Fig.~\ref{fig:ADA-waves}, which justifies why smaller values of $k$ better separate crystals than larger values.
For inorganic crystals, these waves have sharper ``zigzags'' in \cite[Fig.~3]{widdowson2025geographic}.
%\smallskip

All distance-based invariants do not distinguish mirror images, which motivated us to develop the stronger invariant (called \emph{isoset}) whose completeness was proven under isometry \cite{anosova2021isometry} and then adapted to distinguish between mirror images under rigid motion \cite{anosova2026recognition}. 
All invariants and metrics above are computable in polynomial time in the number $k$ of neighbours and the motif size $m$, for a fixed dimension $n$.
\smallskip

A limitation is an approximate algorithm for a metric $d$ between complete isosets.
This algorithm has an approximation factor of about 4 in $\R^3$ \cite[Corollary~5.10]{anosova2026recognition} and guarantees that if the approximation of $d$ is 0, then so is $d$, and hence the underlying crystals are rigidly equivalent.
Similarly, if an approximation of $d$ is small, then $d$ can be at most 4 times larger, which suffices for detecting near-duplicates in practice.
\smallskip

\begin{figure}[h!]
\centering
\includegraphics[width=\textwidth]{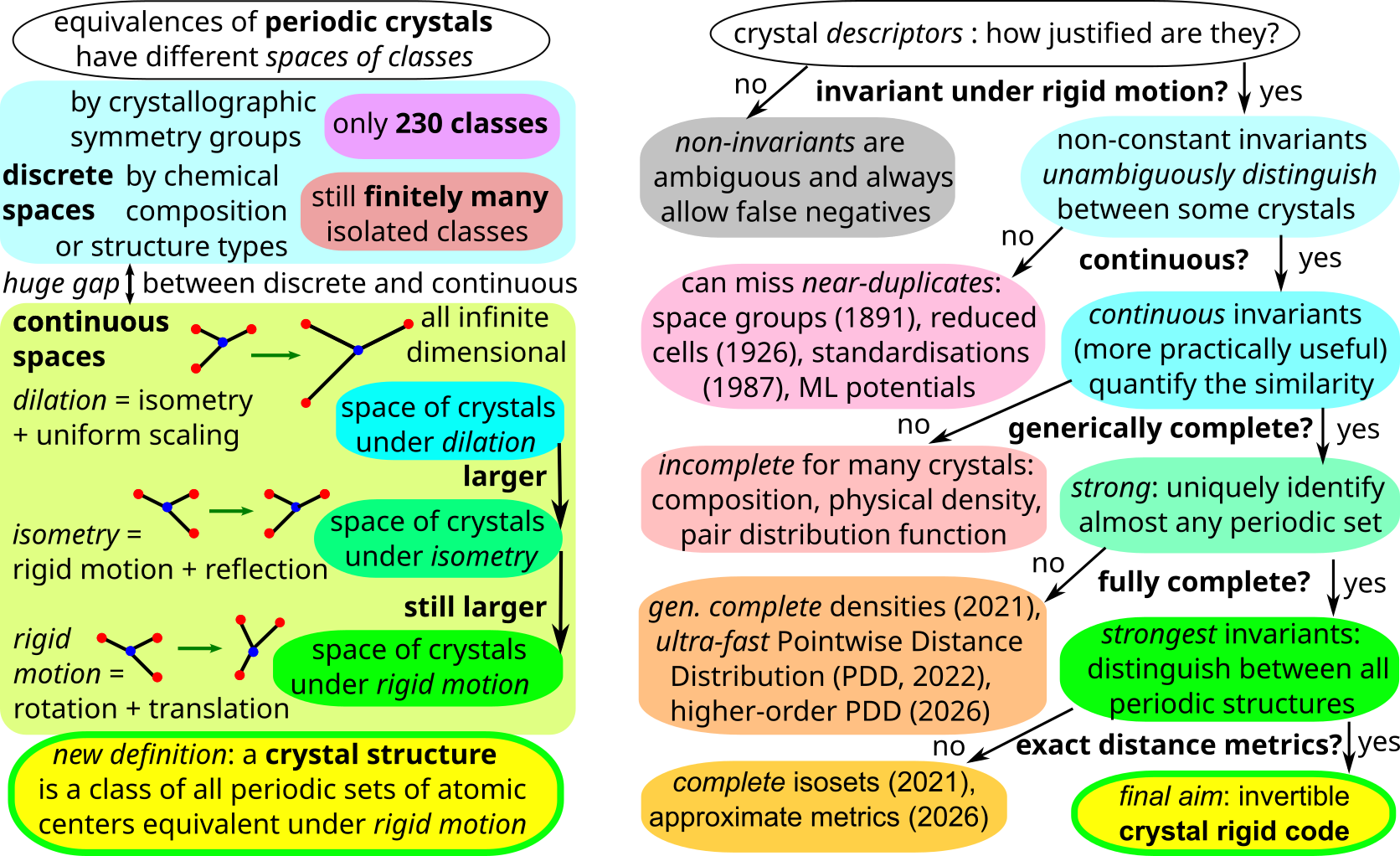}
%\vspace*{-4mm}
\caption{\textbf{Left}: periodic crystals were previously studied by discrete equivalences but are experimentally determined as rigid structures that live in large \emph{continuous spaces}.
\textbf{Right}: the final aim of developing structural invariants is a \emph{crystal rigid code} that is theoretically complete, continuous under tiny perturbations, and invertible back to a periodic crystal structure $S\subset\R^3$, uniquely under rigid motion.}
\label{fig:spaces-invariants}
\end{figure}

Fig.~\ref{fig:spaces-invariants} summarises progress from discrete to continuous classifications, including generically complete density functions \cite{edelsbrunner2021density}, see historical details in \cite[section~12.4]{anosova2025geometric}.
\smallskip

Algorithm~\ref{alg:crystal_neighbours} describes how a crystal structure $S$ given by a CIF is uniquely identified by a hierarchy of invariants.
This identification can involve finding the nearest neighbours of $S$ in a given dataset $D$, such as the CSD \cite{groom2016cambridge}, or several datasets.
Stages 1-4 in Algorithm~\ref{alg:crystal_neighbours}
 are justified by \cite[Theorem 4.4]{widdowson2026pointwise} and \cite[Theorem 6.5]{anosova2026recognition}: a metric between weaker invariants is a lower bound for a metric between stronger ones.

\begin{algorithm}[ht]
\caption{Nearest neighbours of a periodic crystal $S$ in a known dataset $D$}
\textbf{Stage 1}. 
$\ADA(S;k)$ and $\PDA(S;k)$ are computed in near-linear time of $k,m$ \cite[Theorem 3.10]{widdowson2026pointwise}.
If the vectors $\ADA(Q;k)$ are pre-computed for all periodic crystals $Q\in D$, they can be organised in a k-d tree \cite{gieseke2014buffer}, which speeds up the search for a fraction (say, 1\%) of the nearest neighbours to $\ADA(S;k)$ by the metric $\RMS$.
\smallskip

\noindent
\textbf{Stage 2}.
For the nearest neighbours found by $\RMS$ on $\ADA(S;k)$, we compute the EMD on $\PDA(S;k)$.
The $\EMD$ is exactly computed in time $O(m^3\log m)$, where $m$ is the maximum size $\PDA$ distributions \cite{orlin1993faster}, and is approximated in linear time \cite{shirdhonkar2008approximate}.
\smallskip

\noindent
\textbf{Stage 3}.
For a fraction of the nearest neighbours found by $\EMD$ on $\PDA(S;k)$, we compute the $\EMD$ adapted to higher-order $\PDAh(S;k)$, usually only for $h=2$, to distinguish between potential homometric crystals that have equal $\PDA$s.
\smallskip

\noindent
\textbf{Stage 4}.
For a fraction of the nearest neighbours found by $\EMD$ on $\PDAh(S;k)$, we compute the complete isoset, % $I(S;\al)$,  
%depends on a cut-off radius $\al$ and 
which is needed only in rare cases to distinguish between mirror images; there are four such pairs in the whole CSD, see \cite[Fig.~7~(right)]{anosova2026recognition}.
\label{alg:crystal_neighbours}
\end{algorithm}
\smallskip

The invariant $\AMD(S;k)$ practically distinguishes all non-duplicate crystal structures in the CSD already for $k=100$ \cite{widdowson2022average}.
In other words, if any periodic crystals $S,Q$ in the CSD have close $\AMD(S;100)$ and $\AMD(Q;100)$, for example, by $\RMS$ up to $10^{-10}\angstrom$, then they can be matched by isometry with a similar small perturbation.
Here, one Angstrom ($1\angstrom=10^{-10}$ meter) is roughly the smallest inter-atomic distance.
\smallskip

Stages 1-3 sufficed to find the nearest neighbours of all experimental crystals from \cite{szymanski2023autonomous,zeni2025generative} in the Inorganic Crystal Structure Database (ICSD) and Materials Project (MP), including those that were missed in \cite{leeman2024challenges}, see \cite[Table~2]{widdowson2025geographic} and \cite[Tables 1-4]{widdowson2026pointwise}. 
\smallskip

After a news story \cite{chawla2025duplicate} in December 2025 highlighted our research since 2021, the A-lab paper \cite{szymanski2023autonomous} was corrected in January 2026 by crossing out almost all the words ``novelty'' and ``discovery'', and acknowledging that ``the original claims of material novelty were subject to misinterpretation - their intention was to indicate that the materials were new to the prediction platform, not necessarily new to science'' \cite{szymanski2026correction}. 
\smallskip 

The next section shows that our invariants 
%based on cells or motifs can easily miss structural near-duplicates.
scale up to all-vs-all comparisons in the world's largest databases, using Google's GNoME as a high-profile example \cite{merchant2023scaling}. 
%However, a complete hierarchy of continuous invariants can quickly quantify the structural novelty of any newly reported crystal and also filter out numerous near-duplicates and expose physically unrealistic structures in large materials databases

%3====================
\section{A mathematical ``anatomy'' of the GNoME data}
\label{sec:GNoMEs}

This section applies the mathematical tools of section~\ref{sec:definitions} to the GNoME data from \cite{gnome2023data}.
We focus on the first version (called here GNoME1, dated 29 November 2023), which was published on the same day as the Nature paper \cite{merchant2023scaling}, and the last version (called here GNoME2, dated 21 November 2024), see details of all versions in appendix~\ref{sec:GNoME_details}.
\smallskip

The main authors of \cite{merchant2023scaling}, Amil Merchant and Ekin Dogus Cubuk, who also co-authored the A-lab paper \cite{szymanski2023autonomous}, claimed
``the discovery of 2.2 million new crystals – equivalent to nearly 800 years’ worth of knowledge'', which is quoted from \cite{merchant2023millions}.
\smallskip

While 2.2 million ``new crystals'' were not publicly shared, GNoME1 has 384,938 ``most stable'' crystals.
GNoME2 has 554,054 entries, of which 301,413 are shared with GNoME1.
Review \cite[p.~3492]{cheetham2024artificial} mentioned that up to 42 thousands entries in GNoME1 contain radioactive elements \ce{Ac}, \ce{Np}, \ce{Pa}, \ce{Pm}, \ce{Pu}, and \ce{Tc}, but missed Bismuth \ce{Bi} (weakly radioactive), Thorium \ce{Th}, and even Uranium \ce{U}, which we count in Table~\ref{tab:GNoMEs-radioactive}.
 
\begin{table} 
\begin{tabular}{l|rrr|rr}
number of entries with & 1 radioactive element & 2 elements & 3  elements & total & rate \\
\hline
GNoME\textbf{1} & 75,628 & 4,502 & 105 & 80,235 & 20.8\% \\
GNoME\textbf{2} &  102,217 & 5,664 & 139 & 108,020 & 19.5\% \\
\end{tabular} 
\caption{Entries in GNoMEs 1\&2 with radioactive elements. Full tables can be shared by request.}
\label{tab:GNoMEs-radioactive}
\end{table}

%3.1--------------------------------------------
\subsection{Comparing predictions in GNoMEs 1\&2 as textual  CIFs}

%We start by comparing CIFs in the GNoME, without using crystal invariants yet.
For GNoME1, \cite[Table~1]{anosova2024importance} had one quadruple (with IDs 4135ff7bc7, 6370e8cf86, c6afea2d8e, e1ea534c2c), 43 triples, and 1,089 pairs of CIFs that are identical texts symbol-by-symbol, which gives $1\times 4+3\times 43+2\times 1,089=2,311$ entries in total, though \cite{merchant2023scaling} claimed to
``take the top 100 structures for any
given composition and perform pairwise comparisons with pymatgen’s
built-in structure matcher".
\smallskip

When comparing only numerical data without chemical elements, the total number of duplicates rises to 3,248 entries, including a group of 9 CIFs in \cite[Table~2]{anosova2024importance}, where some atoms were replaced without changing their positions.
These replacements affect atomic interactions and should perturb the geometry at least slightly, so keeping all numbers in CIFs to six decimal places in Angstroms seems physically unrealistic. 
%\smallskip

\begin{table}[h!]
\begin{tabular}{L{22mm}|R{15mm}|R{17mm}|R{17mm}|R{17mm}|R{17mm}} 
group size: \textbf{same} CIF geometry &
\textbf{no rounding} &
rounding to \textbf{4 digits} &
rounding to \textbf{3 digits} & 
rounding to \textbf{2 digits} &
rounding to \textbf{1 digit} \\
 \hline
 10 or more & 0 & 0 & 0 & 0 & 97 \\ 
 %24+12+16+6+5+5+5+4+6+2+4+2+6
 9   & 1 & 1 & 1 & 1 & 37 \\
 8   & 0 & 0 & 0 & 0 & 56 \\ 
 7   & 1 & 1 & 1 & 1 & 107 \\
 6   & 2 & 2 & 2 & 3 & 155 \\
 5   & 2 & 3 & 3 & 9 & 304 \\
 4   & 8 & 12 & 13 & 21 & 743 \\
 3   & 72 & 100 &  113 & 181 & 2,447 \\
 2   & 1,481 & 1,994 & 2,141 & 2,633 & 14,274 \\
\hline
all entries & 3,248 & 4,379 & 4,716 & 5,972 & 44,238 \\
percentage  & 0.84\% & 1.14\% & 1.23\% & 1.55\% & 11.49\%
\end{tabular}
\caption{Groups of near-duplicate CIFs in \textbf{GNoME1} are obtained by rounding cell parameters (3 lengths and 3 side diagonals) and Euclidean coordinates of atoms in given CIFs, all in Angstroms.}
\label{tab:GNoME1-near-duplicates}
\end{table}

This intuition was inspired by R.~Feynman in \cite[Fig.1-7]{feynman2011lectures} and stated as the \emph{Crystal Isometry Principle} \cite{widdowson2022average}: under fixed ambient conditions, any real periodic crystal is uniquely determined by a precise enough geometry of only atomic centres without chemical elements, now verified on the experimental databases \cite[section 12.4]{anosova2025geometric}.
\smallskip

Since atoms have fractional coordinates %(usually between 0 and 1) 
in CIFs, it is fairer to compare atomic positions after
converting them into Euclidean coordinates, depending on a given cell basis.
For GNoME1, Table~\ref{tab:GNoME1-near-duplicates} reports groups of CIFs with duplicated geometry after rounding Euclidean coordinates of atoms to $0.0001\angstrom$, $0.001\angstrom$, $0.01\angstrom$, and $0.1\angstrom$, where even the last tolerance is considered acceptable for noisy experimental data \cite{allen1995systematic}. 
\smallskip

\begin{table}[h!]
\begin{tabular}{L{22mm}|R{15mm}|R{17mm}|L{17mm}|L{17mm}|L{17mm}} 
group size: \textbf{same} CIF geometry &
\textbf{no rounding} &
rounding to \textbf{4 digits} &
rounding to \textbf{3 digits} & 
rounding to \textbf{2 digits} &
rounding to \textbf{1 digit} \\
 \hline
 8 or more  & 0 & 0 & 0 & 0 & 347 \\ 
%112+78+27+24+27+21+11+6+8+5+8+3+1+4+1+2+1+1+1+2+1+1+1+1
 7  & 1 & 1 & 1 & 1 & 181  \\
 6   & 3 & 3 & 3 & 3 & 294 \\
 5   & 1 & 2 & 2 & 4 & 613 \\
 4   & 4 & 6 & 9 & 13 & 1,287 \\
 3   & 17 & 55 & 61 & 90 & 3,660 \\
 2   & 247 & 491 & 534 & 858 & 16,947 \\
\hline
all entries & 591 & 1,206 & 1,322 & 2,083 & 60,012 \\
percentage  & 0.11\% & 0.22\% & 0.24\% & 0.38\% & 10.83\%
\end{tabular}
\caption{Groups of near-duplicate CIFs in \textbf{GNoME2} with rounded cell parameters (3 lengths and 3 side diagonals) and Euclidean coordinates of atoms in given CIFs, see groups of 7 and 6 in Table~\ref{tab:GNoME2groupsCIFs}.
}
\label{tab:GNoME2-near-duplicates}
\end{table}

In columns 2-5 of Table~\ref{tab:GNoME1-near-duplicates}, the largest group of 9 CIFs is the same as in \cite[Table~2]{anosova2024importance}.
Table~\ref{tab:GNoME2groupsCIFs} shows two largest groups with identical geometry in CIFs of GNoME2. 

\begin{table}[h!]
\begin{tabular}{l|l||l|l}
GNoME2 id & composition, group of 7 & GNoME2 id &  composition, group of 6 \\
\hline
21a66e0029 & \ce{Cd6 Os1 Ru1 Tb2 Y18} &
02c4cb55a6 & \ce{Cd6 Dy15 Ru2 Tb5 } \\
28566d9d7c & \ce{Cd6 Dy1 Os1 Ru1 Y19} &
0affe9c149 & \ce{Cd6 Dy18 Ru2 Tb2} \\
511b93a2bd & \ce{Cd6 Os1 Ru1 Y20} &
100cfdfdef & \ce{Cd6 Dy17 Ru2 Tb3} \\
7f9b693d17 & \ce{Cd6 Os1 Ru1 Tb1 Y19} &
877c190805 & \ce{Cd6 Dy16 Ru2 Tb4 } \\
885c00d261 & \ce{Dy2 Cd6 Os1 Ru1 Y18 } &
9ce48821cb & \ce{Cd6 Dy20 Ru2} \\
8dfede557b & \ce{Cd6 Os1 Ru1 Tb15 Y5} &
b9e4b78276 & \ce{Cd6 Dy19 Ru2 Tb1} \\
d711c9e131 & \ce{Cd6 Ho1 Os1 Ru1 Y19} & &
\end{tabular}
\caption{Columns 1-2 and 3-4 show the largest groups of 7 and 6 CIFs, respectively, in \textbf{GNoME2} from Table~\ref{tab:GNoME2-near-duplicates} with all identical numbers (6+ decimal places) in their CIFs. }
\label{tab:GNoME2groupsCIFs}
\end{table}

%3.2--------------------------------------------
\subsection{Comparing GNoMEs 1\&2 by a hierarchy of invariants}

The notations $\ADA[k],\PDA[k]$ with a number $k$ of atomic neighbours and $\IS[\al]$ with a radius $\al$ denote the relevant invariants as functions on all crystals.
We will compare all crystals in GNoMEs 1\&2 by Algorithm~\ref{alg:crystal_neighbours} using 
%a series of invariants starting with 
(1) RMS on $\ADA[100]$, then
(2) EMD on $\PDA[100]$, then 
(3) $\EMD$ on $\PDAsq[100]$, finally 
(4) $\EMD$ on $\IS[5]$, where the radius $\alpha = 5\angstrom$ suffices to include the nearest atomic neighbours.  
\smallskip

For a threshold $\tau$, a pair of periodic crystals is called a \emph{geometric match} if all four metrics between their invariants are within $\tau$.
A geometric match is called \emph{geochemical} if the optimal matching of atoms in the computation of EMD on isosets preserves all chemical elements. 
Matched entries form \emph{groups} that are defined as connected components of a graph whose vertices are crystals, while edges join matched entries. 
\smallskip

Tables~\ref{tab:GNoME1_duplication}-\ref{tab:GNoME2_duplication} count groups of geometric and geochemical matches in GNoMEs 1\&2 for various thresholds $\tau$.
The statistics of duplication is summarised by the \emph{excess} counting entries that would have to be deleted to leave one representative per group.
The \emph{redundancy} rate is the excess divided by the size of the dataset.
Appendix~\ref{sec:GNoME_details} contains more details of the computed invariants, metrics, and their running times.
\smallskip

\begin{table}[htbp]
  \centering
\begin{tabular}{lrrrrrr}
threshold & number of groups of & groups of & entries in & excess & redundancy & largest \\
    
$\tau$, $\angstrom$  & \textbf{geometric} matches & size$\geq 2$ & groups $\geq 2$ & entries & rate, \% & group \\    \hline
    %\multicolumn{7}{l}{\emph{geometric matches}} \\
    $10^{-10}$ & 383,257 & 1,567  & 3,248  & 1,681  & 0.437  & 9 \\
%    $10^{-5}$  & 382,609 & 2,167  & 4,496  & 2,329  & 0.605  & 9 \\
    $10^{-4}$  & 382,446 & 2,324  & 4,816  & 2,492  & 0.647  & 9 \\
    $10^{-3}$  & 381,908 & 2,744  & 5,774  & 3,030  & 0.787  & 9 \\
    $10^{-2}$  & 337,060 & 19,610 & 67,488 & 47,878 & 12.438 & 352 \\
    \hline
    %\multicolumn{7}{l}{\emph{geometric\,$+$\,chemical}} \\
 & \textbf{geochemical} matches & & & & & \\
    $10^{-10}$ & 383,760 & 1,133 & 2,311 & 1,178 & 0.306 & 4 \\
%    $10^{-5}$  & 383,614 & 1,274 & 2,598 & 1,324 & 0.344 & 4 \\
    $10^{-4}$  & 383,468 & 1,417 & 2,887 & 1,470 & 0.382 & 4 \\
    $10^{-3}$  & 383,250 & 1,618 & 3,306 & 1,688 & 0.439 & 4 \\
    $10^{-2}$  & 380,488 & 4,172 & 8,622 & 4,450 & 1.156 & 12 \\
  \end{tabular}
  \caption{Near-duplicates, excess, and redundancy of the \textbf{GNoME1} dataset (384,938 entries). }
\label{tab:GNoME1_duplication}
\end{table}

\begin{table}[htbp]
  \centering
  \begin{tabular}{lrrrrrr}
threshold & number of groups of & groups of & entries in & excess & redundancy & largest \\
    
$\tau$, $\angstrom$  & \textbf{geometric} matches & size$\geq 2$ & groups $\geq 2$ & entries & rate, \% & group \\
\hline
$10^{-10}$ & 553,736 & 273    & 591     & 318    & 0.057  & 7 \\
%    $10^{-5}$  & 553,355 & 599    & 1,298   & 699    & 0.126  & 7 \\
    $10^{-4}$  & 553,339 & 613    & 1,328   & 715    & 0.129  & 7 \\
    $10^{-3}$  & 552,905 & 958    & 2,107   & 1,149  & 0.207  & 7 \\
    $10^{-2}$  & 469,962 & 24,194 & 108,286 & 84,092 & 15.178 & 723 
    %\\     \hline    \multicolumn{7}{l}{\emph{geometric\,$+$\,chemical}} \\
%    $10^{-10}$ & 554,054 & 0 & 0 & 0 & 0.000 & 1 \\
%    $10^{-5}$  & 554,054 & 0 & 0 & 0 & 0.000 & 1 \\
%    $10^{-4}$  & 554,054 & 0 & 0 & 0 & 0.000 & 1 \\
%    $10^{-3}$  & 554,054 & 0 & 0 & 0 & 0.000 & 1 \\
%    $10^{-2}$  & 554,054 & 0 & 0 & 0 & 0.000 & 1 \\
  \end{tabular}
  \caption{Near-duplicates, excess, and redundancy of the \textbf{GNoME2} dataset (554,054 entries). No pairs of entries of GNoME2 were a geochemical match for thresholds up to $0.01\angstrom$. }
\label{tab:GNoME2_duplication}
\end{table}

In comparisons, we used the entries in the ICSD and MP \cite{MP2021} %with the parameter ``include\_gnome=False'', 
only up to March 2021 because ``all candidates for GNoME are derived from snapshots of databases
made in March 2021, including the Materials Project" \cite{merchant2023scaling},
%because the MP started to include some crystals from the GNoMEs, 
see details in appendix~\ref{sub:GNoMEs-vs-ICSD-MP}.
\smallskip

The top 4-row block of Table~\ref{tab:GNoMEs-vs-ICSD_MP_distances} compares GNoMEs 1\&2 with the ICSD and MP using the EMD on invariants PDA[100], which distinguished all non-isometric crystals in the CSD and ICSD.
The bottom 4-row block of Table~\ref{tab:GNoMEs-vs-ICSD_MP_distances} reports the Root Mean Square Deviation (rmsd\_max) by StructureMatcher with default parameters \cite{ong2013python}.
This RMSD measures the deviation between all atoms in suitably extended and optimally aligned unit cells from two crystals, including non-rigid averaging of lattices in \cite[step 10.b.i.ia]{pymatgen}.
The last column in Table~\ref{tab:GNoMEs-vs-ICSD_MP_distances} implies that over 60\% of predictions in GNoMEs 1\&2 are geometrically close to 
known crystals in the ICSD and MP up to  $0.05\angstrom$. 

\begin{table}[h!]
  \centering
  \caption{Statistics of distances from GNoMEs 1\&2 to the ICSD and MP, see histograms in Fig.~\ref{fig:GNoMEs_RMSD_ICSD_MP}. }
  \label{tab:GNoMEs-vs-ICSD_MP_distances}
  \setlength{\tabcolsep}{5.5pt}
  \begin{tabular}{ll|rrrrr|rrr}
   GNoME & reference & \multicolumn{5}{c}{\textbf{EMD} on $\PDA[100]$ in Angstroms for quantiles} & \multicolumn{3}{c}{and under a threshold $\tau$} \\

  version & database & 5\% & 10\% & 25\% & 50\% & 75\% & $0.01\angstrom$ & $0.02\angstrom$ & $0.05\angstrom$ \\  \hline

    \textbf{1} & ICSD & 0.0123 & 0.0166 & 0.0252 & 0.0376 & 0.0547 & 3.3\% & 15.2\% & 69.7\% \\

    \textbf{2} & ICSD & 0.0124 & 0.0164 & 0.0241 & 0.0357 & 0.0516 & 3.0\% & 16.3\% & 73.1\% \\
    \hline

    \textbf{1} & MP   & 0.0099 & 0.0139 & 0.0217 & 0.0334 & 0.0502 & 5.1\% & 21.3\% & 74.8\% \\

    \textbf{2} & MP   & 0.0096 & 0.0131 & 0.0204 & 0.0311 & 0.0466 & 5.5\% & 24.1\% & 78.4\% \\

\hline \hline

   &  & \multicolumn{5}{c}{\textbf{RMSD} by StructureMatcher in $\angstrom$ for quantiles} & \multicolumn{3}{c}{and under a threshold $\tau$} \\  \hline

    \textbf{1} & ICSD & 0.0069 & 0.0106 & 0.0188 & 0.0312 & 0.0484 & 7.4\% & 22.6\% & 62.8\% \\

    \textbf{2} & ICSD & 0.0074 & 0.0108 & 0.0183 & 0.0301 & 0.0468 & 7.3\% & 24.2\% & 66.1\% \\
    \hline

    \textbf{1} & MP   & 0.0057 & 0.0088 & 0.0158 & 0.0275 & 0.0446 & 10.8\% & 30.6\% & 71.1\% \\

    \textbf{2} & MP   & 0.0058 & 0.0086 & 0.0148 & 0.0258 & 0.0424 & 11.8\% & 34.1\% & 74.3\%
  \end{tabular}

\end{table}
\begin{figure}[h!]
\centering
\includegraphics[height=45mm]{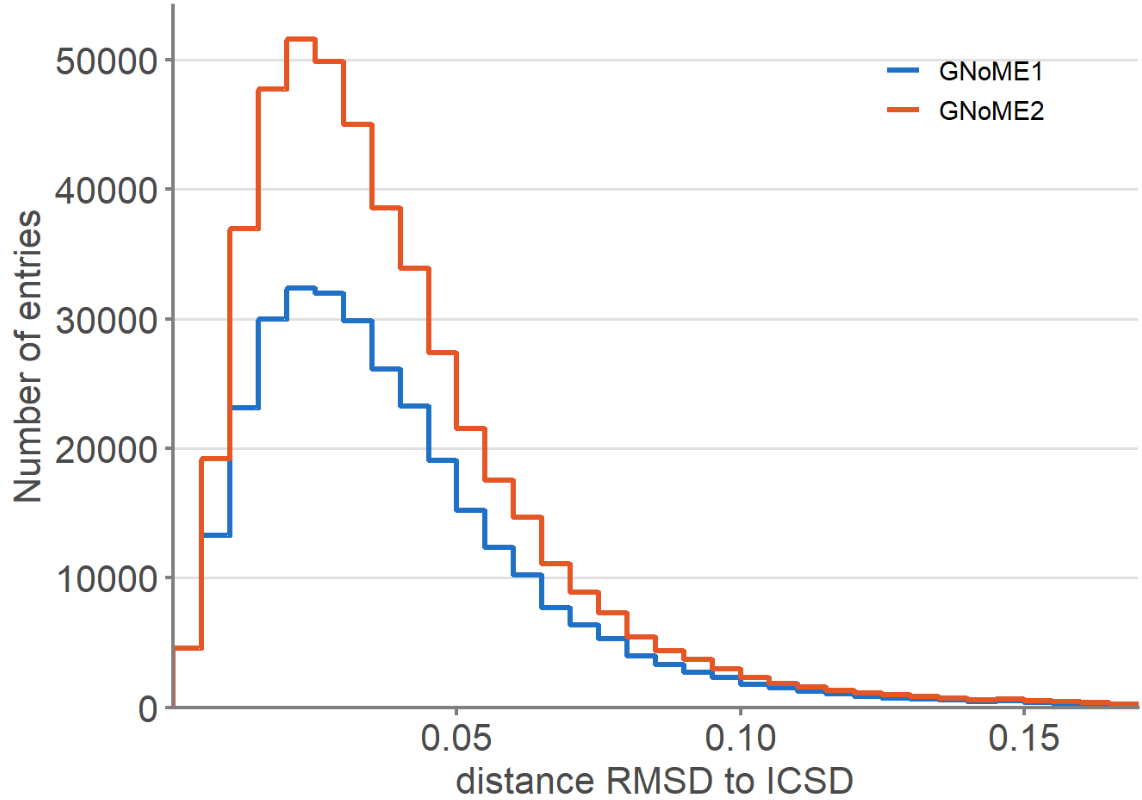}
\includegraphics[height=45mm]{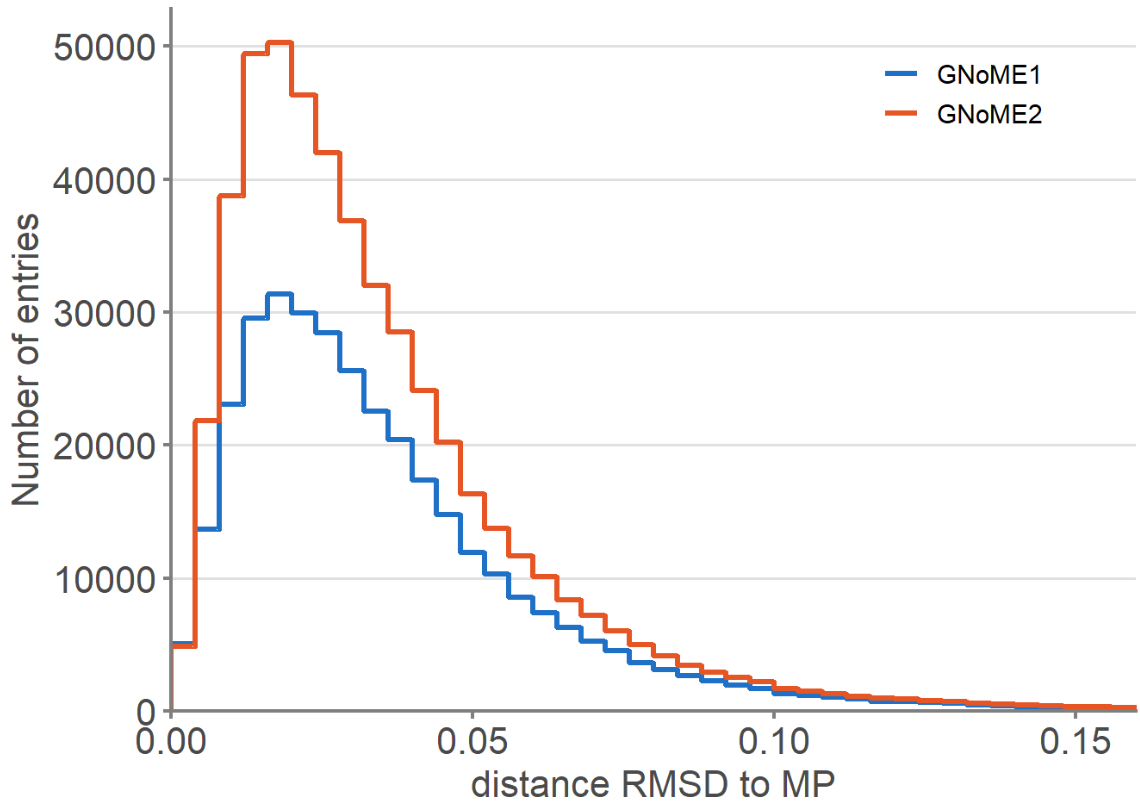}
\caption{The histograms of RMSD in Angstroms, obtained by StructureMatcher in pymatgen \cite{ong2013python} between crystals in GNoMEs 1\&2 and their neighbours in known databases. 
\textbf{Left}: ICSD.
\textbf{Right}: MP.
}
\label{fig:GNoMEs_RMSD_ICSD_MP}
\end{figure}

%3.3----------------------------------
\subsection{Exposing the evolution of GNoME1 into GNoME2}

%\emph{Duplication within each release fell sharply, and chemical duplicates were eliminated outright}. 
Comparisons by crystal invariants reveal that GNoME1 largely evolved into GNoME2 by removing all groups of \emph{geochemical matches} (with identical geometry and chemistry) by deleting full groups rather than by keeping one representative.
In GNoME1 at the threshold $\tau = 10^{-10}\angstrom$ (close to a floating-point error), of the 1,133 groups of geochemical matches, 1,010 groups (89\%) disappeared and only 123 collapsed to a single entry in GNoME2. 
Loosening the threshold reverses the result: at $\tau = 10^{-2}\angstrom$ only 1,817 of 4,172 groups (44\%) were deleted, while 2,355 groups (56\%) shrank. 
\smallskip

GNoME2 introduced extra geometric duplication. % in comparison with GNoME2. 
Of 273 groups of geometric duplicates in GNoME2 at $\tau = 10^{-10}\angstrom$, 179 groups were duplicated in GNoME1, but 62 grew out of GNoME1 singletons, and 32 consist entirely of newly added entries. 
At $10^{-2}\angstrom$, the corresponding counts are 12,979, 5,047, and 6,168 of 24,194 groups. 
%\smallskip

\begin{table}[h!]
  \centering
  \begin{tabular}{lrrrrrr}
%    & & & & partial & & \\
    $\tau, \angstrom$  & \textbf{geometric} groups & deleted & partly deduplicated & deduplicated & kept & grown \\ \hline
    % \multirow{5}{*}{geom.\}
 %   \multicolumn{7}{l}{\emph{geometric}} \\
      $10^{-10}$ & 1,567  & 1,064 & 324   & 12  & 160   & 7 \\
      $10^{-5}$  & 2,167  & 1,170 & 609   & 23  & 349   & 16 \\
      $10^{-4}$  & 2,324  & 1,239 & 689   & 23  & 357   & 16 \\
      $10^{-3}$  & 2,744  & 1,342 & 832   & 45  & 492   & 33 \\
      $10^{-2}$  & 19,610 & 2,693 & 3,938 & 845 & 7,494 & 4,640 \\
    \hline
%    \multicolumn{7}{l}{\emph{geometric\,$+$\,chemical}} \\
    & \textbf{geochemical} groups & & & & & \\
      $10^{-10}$ & 1,133 & 1,010 & 123   & 0 & 0 & 0 \\
      $10^{-5}$  & 1,274 & 1,083 & 191   & 0 & 0 & 0 \\
      $10^{-4}$  & 1,417 & 1,147 & 270   & 0 & 0 & 0 \\
      $10^{-3}$  & 1,618 & 1,234 & 384   & 0 & 0 & 0 \\
      $10^{-2}$  & 4,172 & 1,817 & 2,355 & 0 & 0 & 0 \\
  \end{tabular}
  \caption{What became of GNoME1 matched groups in GNoME2.  
Only groups of sizes $s_1\geq2$ in GNoME1 are listed.
  \emph{Deleted}: every member dropped (new size $s_2 = 0$);
  \emph{deduplicated}: exactly one member survives ($s_2 = 1$);
  \emph{partially deduplicated}: $1 < s_2 < s_1$;
  \emph{kept}: $s_2 = s_1 > 1$;
  \emph{grown}: $s_2 > s_1$.}
  \label{tab:GNoME1_groups_fates}  
\end{table}

More than half of geometric redundancy in GNoME2 is inherited from GNoME1, i.e. 179 of 273 groups at $10^{-10}\angstrom$ and 12,979 of 24,194 at $10^{-2}\angstrom$.
Only a minority (32 groups at $10^{-10}\angstrom$ and 6,168 at $10^{-2}\angstrom$) consists entirely of newly added entries.
Tables~\ref{tab:GNoME1_groups_fates} and \ref{tab:GNoME1_entries_fates} summarise the fates of groups and matched entries, respectively, when GNoME1 evolved into GNoME2.
Fig.~\ref{fig:GNoME1_groups_fates} visualises the fates at several thresholds.

%3.4----------------------------------
\subsection{The most striking examples in the GNoMEs 1\&2 datasets}

%\emph{The most unusual and physically impossible structures in GNoME1 were removed entirely}. 

In GNoME1, among 2,311 entries with repeated CIFs, the entries cdc06a1a2a and 0e2d8f26d6 stand out because each CIF has two pairs of repeated atoms \ce{Na} at the same position in \cite[appendix~A, Fig.~15]{anosova2026recognition} and is not parsable by pymatgen \cite{pymatgen}.
\smallskip

\begin{figure}[h!]
\centering
\includegraphics[height=33mm]{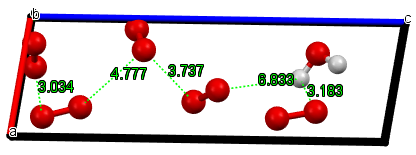}
\includegraphics[height=33mm]{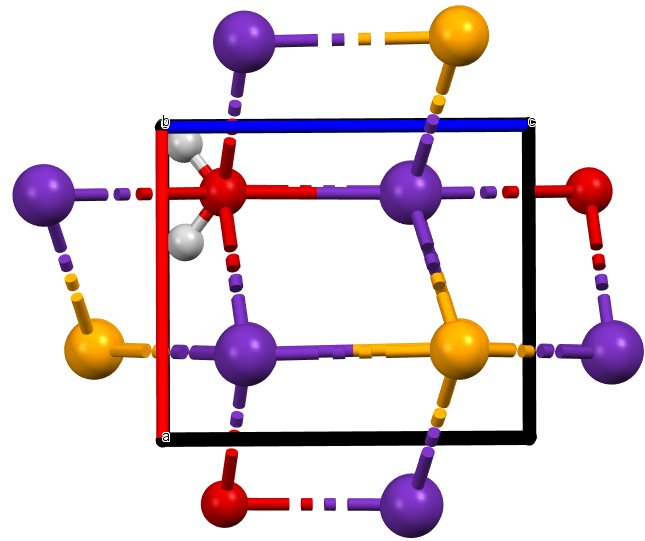}
\caption{
\textbf{Left}: entry 83cc15ef46 (\ce{H2O11}) in GNoME1. 
\textbf{Right}: entry 8175bee583 (\ce{H2ORb2Se}) in both GNoMEs has a water molecule \ce{H2O} in the top left corner, all bonds are visualised by Mercury.}
\label{fig:GNoMEs_water}
\end{figure}

In GNoME1, entry 83cc15ef46 (\ce{H2O11}) consists of chaotically located molecules of water \ce{H2O} and oxygen \ce{O2} in ratio $1:5$, and is unlikely to be a new useful material.
While the authors of \cite{merchant2023scaling} aimed to ``introduce novel compositions in the dataset instead of similar structures that may be missed by structure matchers",
Fig.~\ref{fig:GNoMEs_water} shows that ``new compositions'' can be obtained from existing ones by just adding water (ice).
\smallskip

To further identify unusual entries, we used 
%the four invariants of a periodic crystal $S$ below.
the smallest and largest distances to the first atomic neighbour, denoted by $\dmin$ and $\dmax$, respectively. % are the minimum and maximum values in $\PDD(S;1)$.
The point density $\rho=\dfrac{m}{\vol[U]}$ is the number $m$ of motif points divided by the volume of the unit cell, given in atoms per nm$^3$.
The physical density is %the mass of atoms in a unit cell $U$ divided by $\vol[U]$ and 
measured in g/cm$^3$. 
%\smallskip

\begin{figure}[h!]
\centering
\includegraphics[height=37mm]{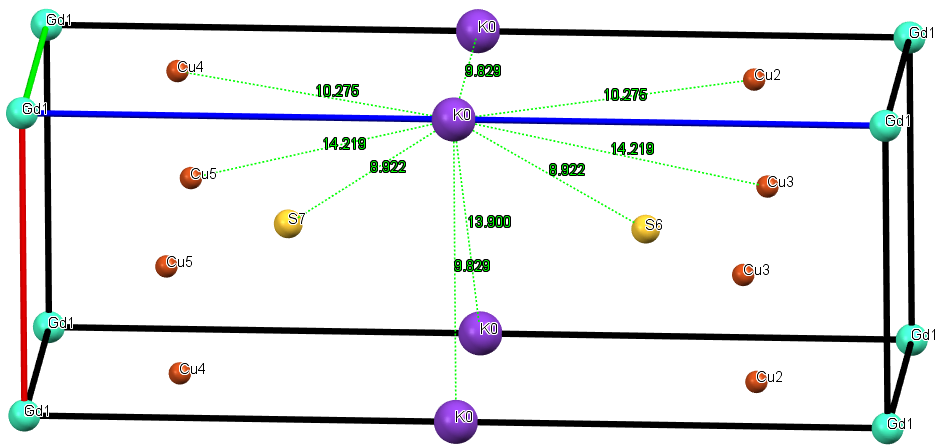}
\includegraphics[height=37mm]{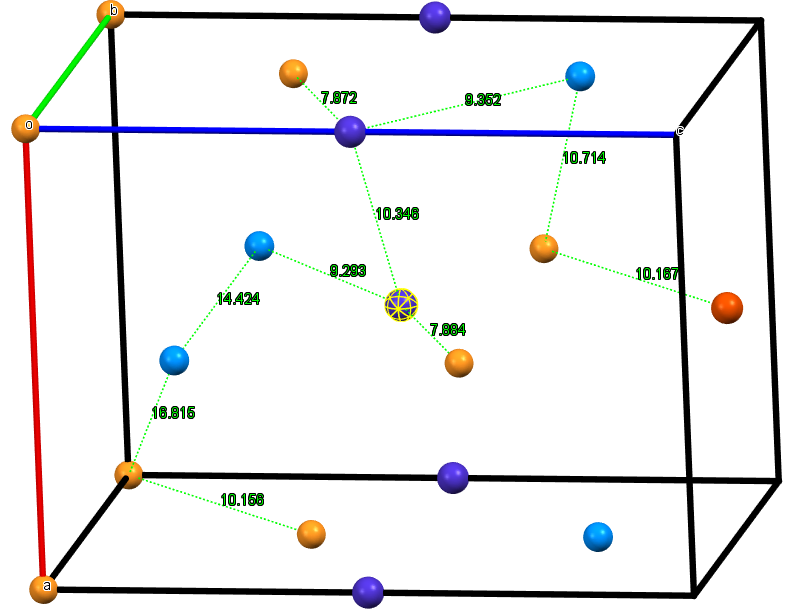}
\caption{
Unrealistic predictions in GNoME1.
\textbf{Left}: entry eed3027c1e has atom \ce{K} with no neighbours up to $\dmax=8.92\angstrom$. 
\textbf{Right}: entry 37f9f79871 has no pairs of atomic neighbours up to $\dmin=7.87\angstrom$.
}
\label{fig:GNoME1_largest_dmax_dmin}
\end{figure}

Significant deviations of these basic invariants from typical values in experimental datasets indicate that a crystal may not be physically realistic. 
In GNoME1, the sparsest entry 88e5e5c7a3 (\ce{S7Gd2Tb3}) has $1.29$ atoms per nm$^3$, forty times fewer than the median value in GNoME1, with a density of 0.18 g/cm$^3$. 
Entry eed3027c1e (\ce{KGd(Cu2S)2}) has the largest $\dmax=8.92\angstrom$ with a \ce{K} atom floating too far away from all others. 
Entry 37f9f79871 (\ce{P3FeCo2U2}) has $1.81$ atoms/nm$^3$, density $0.28$ g/cm$^3$, and $\dmin = 7.87\angstrom$, so atoms in this prediction have no close contact at all, see Fig.~\ref{fig:GNoME1_largest_dmax_dmin}. 

%4====================
\section{Scaling easy disguises of real crystal structures}
\label{sec:disguises}
Atomic vibrations and experimental noise inevitably produce slightly different cell parameters and atomic coordinates, whether that is due to different instruments, software, or conditions, such as temperature or pressure.
A manual search for near-duplicates of 43 A-lab crystals in the ICSD by unit cell parameters missed several crystals that appeared in the ICSD earlier with different unit cells, see \cite[Fig.~5]{widdowson2025geographic}. 
\smallskip

The discontinuity of reduced cells under trivial perturbations, as in Fig.~\ref{fig:basis_ambiguity}~(top right), implies that any comparison based on a unit cell or a single (even if large) finite subset can inevitably miss many near-duplicates, as we found with the GNoMEs. % and other databases \cite{widdowson2026pointwise}.
\smallskip

More importantly, the discontinuity of cell-based representations allows anyone to disguise an existing experimental crystal by the following steps:
\smallskip

\noindent
(1) Shuffling (lines of) atoms and their indices in a CIF and shifting the origin. 
\smallskip

\noindent
(2) Applying a linear transformation that changes and scales up a given unit cell.
\smallskip

\noindent
(3) Adding noise to the atom positions to make the new scaled-up cell primitive.
\medskip

While the steps (1)-(2) keep the crystal structure rigidly equivalent to the original, a CIF substantially changes.
Moreover, any raw data, such as diffraction patterns, can also be slightly perturbed.
%This disguise goes beyond CIF-level duplicates in the GNoMEs.
%For a non-P1 space group, a reduction to standard settings might recognise the original crystal in theory, though any computations accumulate floating-point errors, which can give a binary ``no'' answer.
%For the P1 group of only lattice translations, recognising duplicates also requires finding a suitable shift matching atoms within a fixed unit cell.
Step (3) discontinuously affects a primitive cell by size and hence makes all attempts at cell reduction obsolete by disguising a crystal beyond the capacity of cell-based tools, see details of StructureMatcher \cite{pymatgen} in appendix~\ref{sub:GNoMEs-vs-ICSD-MP}.  
\smallskip

The packing similarity in Mercury compares a fixed number of molecules (15 by default) only in organic crystals.
Adapting it to inorganic crystals would require exponentially many matchings of chemically identical atoms, as in the organic case, and still would not guarantee continuity because of a cut-off radius.
For complete isosets with a cut-off, the discontinuity was resolved by a boundary-tolerant metric in \cite[Definition~4.1]{anosova2026recognition} without additional parameters that are used in SOAP descriptors \cite{bartok2013representing}.
\medskip

If novelty is evaluated only by chemical composition, one can replace chemical elements with similar ones.
Hence, it is trivial to generate infinitely many different CIFs without essentially changing the underlying geometry of a crystal structure.
\smallskip

\begin{figure}[h!]
\centering
\includegraphics[width=\textwidth]{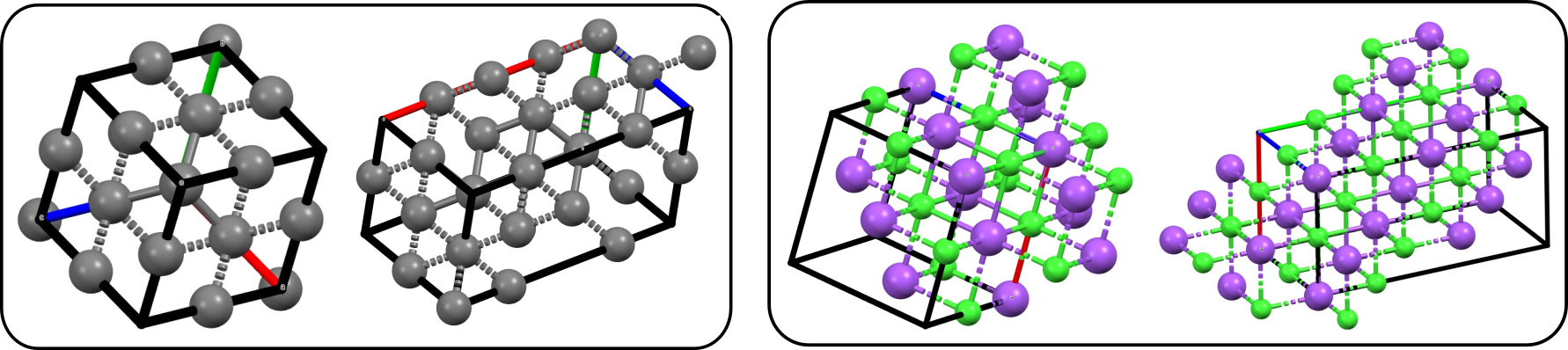}
\vspace*{-2mm}
\caption{An original crystal and its near-duplicate that has a different CIF with a larger primitive cell.
\textbf{Left}: diamond, mp-394 in the Materials Project (MP).
\textbf{Right}: \ce{NaCl}, mp-22862 in the MP.
}
\label{fig:C+NaCl_near-duplicates}
\end{figure}

The second author implemented a white-box algorithm \cite{kurlina2026disguise} that can generate 2.2 million new-looking, realistic crystals in about a day on a small laptop, using transparent school-level geometry instead of black-box AI, see details in appendix~\ref{sec:crystal_disguise}.
\smallskip

The program can disguise any existing crystal so that its CIF looks completely different, and the final crystal is guaranteed to be physically realistic as a near-duplicate of the original, as in Fig.~\ref{fig:C+NaCl_near-duplicates}.
The first presentation of this crystal disguise received a Science Slam award at the European Crystallographic Meeting in August 2024.
%Filters and similarity metrics that do not use Lipschitz continuous metrics cannot reliably recognise the disguised CIF as a geometric near-duplicate.
%Uniform distribution of points on a 2D sphere and in a 3D ball \cite[section~3.29]{fishman2013monte}.

%5====================
\section{Necessary conditions for next-generation real AI}
\label{sec:conditions}

John Maddox \cite{maddox1988crystals} acknowledged the ``general impossibility'' of crystal structure prediction as a ``scandal'' in 1988, and Sally Price FRS \cite{price2018zeroth} later rephrased it as the ``embarrassment of over-prediction'' because very few predictions are synthesised.
\medskip

The far greater ``scandal'' was the absence of a rigorous definition of a crystal structure, which fuelled a ``crisis of trust'' in applied sciences, including paper mills \cite{bimler2022better} and large-scale retractions in crystallography \cite{harrison2010editorial}, which continue to this day \cite{kurlin2026duplicate}.
\medskip

Indeed, simulations can produce numerous near-duplicates as slightly different approximations to the same local minimum of complicated energy \cite{mardirossian2017thirty}.
The past approach was to report ``meta-stable'' predictions close to the boundary of the convex hull on points (composition, energy) within a small threshold like 1 meV/atom \cite{gnome2023data}.
\medskip
 
Any fixed threshold lets us generate infinitely many near-duplicates, all with different CIFs and similar geometries that should have energies close to the current boundary of the convex hull.
For the ``most stable'' crystals exactly on the boundary, some fraction of their trivial perturbations have slightly different energies and likely appear on a ``new'' boundary in a high-dimensional space without predicting new materials. 
Past measures, such as SUN (stability, uniqueness, novelty), used cell-based comparisons but started to include the first continuous invariant $\AMD[k]$ \cite{negishi2026continuous,babu2026toward}.
\medskip

After a (periodic) crystal structure was mathematically defined in June 2024 as a class of all crystal representations that are equivalent under rigid motion \cite[Definition 6]{anosova2024importance} and after the development of a hierarchy of continuous invariants \cite{widdowson2022average,widdowson2026pointwise,widdowson2027higher,anosova2026recognition}, the scale of artificial imitation in materials discovery has become much clearer. 
\medskip

To tackle the artificially created crisis in computational trial-and-error at scale, we propose the following necessary conditions for any next-generation explainable AI:
\medskip

\noindent
(1) \textbf{Input credibility} for periodic crystals should be guaranteed in terms of stable-under-noise invariants instead of discontinuous descriptors.
Indeed, a cell-based CIF of an infinite periodic crystal is like a single snapshot of a complicated rigid structure, which can have infinitely many photographs taken from different points of view. 
\medskip

\noindent
(2) \textbf{Chemical utility} should be evaluated by domain experts beyond comparing compositions as strings.
Also, radioactive and toxic elements should be avoided.
\medskip

\noindent
(3) \textbf{Structural novelty} of any crystal $S$ with respect to a dataset $D$ of known crystals should be quantified by the continuous \emph{Local Novelty Distance} $\LND\langle d, I, D\rangle(S)=\min\limits_{Q\in D} d(I(S),I(Q))$ for a metric $d$ on a continuous invariant $I$ of crystal structures. 
\medskip

\noindent
(4) \textbf{The atomic geometry} should satisfy reasonable constraints by quantifying the average and maximum deviations from expected covalent and ionic bond lengths, depending on chemical elements, to avoid `new' but physically unrealistic predictions. % with too close or too distant atoms.
\medskip

The past work \cite{widdowson2025geographic} introduced $\LND$ using $\EMD$ on the invariant $\PDA[100]$ to quantify the novelty of 43 predictions by the A-lab relative to the ICSD and MP.
This LND has been adopted as the first stress test for AI-guided materials discovery \cite{zhang2026new}.
\medskip

%Our paper extended the $\LND$ to complete isosets, because only complete invariants can confirm duplication without manual inspection. 
%Fig.~\ref{fig:GNoMEs_EMD_PDA100_ICSD_MP} continuously quantifies structural novelty of GNoMEs 1\&2 relative to the ICSD and MP of known crystals.
Continuous structural novelty is needed to reliably filter out near-duplicates %and physically unrealistic structures 
in any database.
Without deduplicating an initial dataset, any splitting into 80:20 subsets for training and testing can keep near-duplicates in each part.
Preventing this \emph{data leakage} %\cite{shabtai2012data}.
requires formalising the basic questions ``Same or different? If different, by how much?'' for all types of data under non-trivial equivalences in \cite[Problem~1.4.5]{anosova2025geometric}.
\smallskip

Conditions (1)-(4) are only minimal requirements for any  justified AI and should 
%evolve into sufficient conditions by including
be complemented in the future with 
more physics-based criteria and experimental evidence, regarding stability (synthesisability) and utility (functional properties).
\smallskip

In particular, there is no sense in maximising any version of $\LND$ without constraints on atomic geometry, see more unrealistic predictions in
Fig.~\ref{fig:GNoME_LND_EMD_PDA100_ICSD} and Table~\ref{tab:GNoME_LND_EMD_PDA100_ICSD}.
After embracing Goodhart's law ``when a measure becomes a target, it ceases to be a good measure''
 \cite{fire2019over}, science can finally progress beyond optimising single-value functions, which are unlikely to characterise the synthesizability of real materials.
\smallskip

Concerning experimental validation of the GNoME,  \cite[p.~29638]{miura2024efficient} said that ``we first search for \ce{Cs}$_n$\ce{ABCl6} structures that are computationally stable or close to stable in the combined zero-temperature convex hull of crystals from Materials Project, OQMD, and GNoME databases'' and then reported successful in situ XRD for materials  \ce{Cs2LiCrCl6}, \ce{Cs2LiFeCl6}, \ce{Cs2LiIrCl6}, \ce{Cs2LiRhCl6}, \ce{Cs2LiRuCl6}, \ce{Cs2NaDyCl6} in \cite[Table~1]{miura2024efficient}.
None of these compositions were found in GNoMEs 1\&2.
The only other compound, synthesised with impurities, was \ce{Cs2LiVCl6}.
For this composition, prediction 66ae3fec06
appeared in GNoME1 (not in GNoME2) and has the following nearest neighbours in the ICSD and MP with small $\EMD$s on invariants $\PDA[100]$:
\smallskip

\noindent
ICSD 002064 (\ce{Cl3CsCu}), published in 1974 \cite{kroese1974high} and deposited in 1980, has two \ce{Cu} atoms, which are replaced with \ce{Li} and \ce{V} to match 66ae3fec06 with $\EMD\approx 0.064\angstrom$.
\smallskip

\noindent
mp-1206187 (\ce{Cl6LiRb2Rh}), published in 1979 \cite{meyer1979zur},  has atoms \ce{Rb} and \ce{Rh}, which are are replaced with \ce{Cs} and \ce{V}, respectively, to match 66ae3fec06 with $\EMD\approx 0.043\angstrom$.
\smallskip

The same paper \cite[p.~29638]{miura2024efficient} added that ``we selected the compounds that did not have a matching ICSD entry (by composition) ... identifying compounds not in the ICSD is commonly considered a benchmark for experimental novelty''.
To excel in this ``novelty'' benchmark, it suffices to just add water (ice), as illustrated in Fig.~\ref{fig:GNoMEs_water}. %, by putting a sample in a fridge.
\medskip

Another paper \cite{naganuma2026experimental} experimentally verified a compound ``\ce{MnFeCo4Si2} discovered by GNoME''.
For this composition, the prediction d8afdc992d in both GNoMEs has a near-duplicate 675046 in the ICSD with $\EMD\approx 0.022\angstrom$, deposited in 2018 \cite{abderrahim2016half}.
\smallskip
 
Any finite dataset is a measure-zero subset of the space of all continuous data, including physical structures whose atoms have real coordinates.
For more practical value, finitely many examples can help formulate conjectures that should be converted into justified guarantees, such as generic completeness and Lipschitz continuity. 
\smallskip

For instance, 100 values of the invariant $\ADA(S;100)$ sufficed to put all non-isometric crystals in major databases at their unique locations on a continuous map of the materials space.
This Crystal Isometry Space, introduced in \cite[Definition~1]{widdowson2025geographic}, extends Mendeleev's table of chemical elements to unambiguously identify all periodic crystals, like houses on a geographic map through infinite-precision satellite cameras.
After that, whether to assign different houses to one village (cluster) or another one using manually chosen thresholds is more an administrative than a scientific decision.
\smallskip

While both humans and AI can make mistakes, meaningful progress means learning from them. 
Just as Isaac Newton transformed the simple mistake of sitting under an apple tree into the law of gravity, we welcome collaborations that advance sustainable science towards rigorously justified solutions based on real Analytic Intelligence.
\medskip

We thank Richard Catlow FRS, Andy Cooper FRS, Sally Price FRS, Simon Billinge, Ram Seshadri, Robert Palgrave, Leslie Schoop, and Mike Glazer for fruitful discussions.
This work was funded by the Royal Society Short Industry Fellowship (SIF/R2/252048) at the Cambridge Crystallographic Data Center, which re-implemented the PDD invariants for inclusion in the release of Mercury in 2027.

\bibliography{necessary-conditions-AI-materials}% common bib file
%\input sn-article.bbl

%\end{document}

%The drafted appendices below will be polished for a public version by October.

\setcounter{section}{0}
\renewcommand{\thesection}{\Alph{section}}

%A======================
\section{Details of mathematics for periodic crystals}
\label{sec:math_details}

%Def------------------------------------
\subsection{Mathematical definitions of crystallographic concepts}
\label{sub:crystallographic_concepts}

We start from a classical representation as in a Crystallographic Information File (CIF) and remind of the new concept of a crystal structure \cite{anosova2024importance} in Definition~\ref{dfn:structures}.
\medskip

\begin{dfn}[lattice, unit cell, periodic crystal] 
\label{dfn:periodic}
\textbf{(a)}
Vectors $\vec v_1,\dots,\vec v_n\in\R^n$ form a \emph{basis} if any vector in $\R^n$ can be written as $\vec v=\sum\limits_{i=1}^n t_i\vec  v_i$ for unique $t_1,\dots,t_n\in\R$. 
\smallskip

\noindent
\textbf{(b)}
Any basis $\vec v_1,\dots,\vec v_n\in\R^n$ defines the \emph{lattice} $\La=\{\sum\limits_{i=1}^n c_i\vec  v_i \mid c_1,\dots,c_n\in\Z\}$ and the \emph{unit cell} $U=\{\sum\limits_{i=1}^n x_i\vec  v_i \mid 0\leq x_i<1,\; i=1,\dots,n\}\subset\R^n$.
\smallskip

\noindent
\textbf{(c)}
For any finite \emph{motif} of points $M\subset U$, the sum $S=M+\La=\{\vec p+\vec v \mid p\in M, v\in\La\}$ is called a \emph{periodic point set}.
In dimension $n=3$, if every point of $S$ has an \emph{atomic type}, which is a chemical element, possibly with a charge and or a mass number (for isotopes of the same element), then $S$ can be called a \emph{periodic crystal}. 
\end{dfn}
\medskip

A periodic crystal $S\subset\R^3$ can generate many periodic point sets.
We can take points at all atomic centres of $S$, or points only at centres of specific atoms, or centres of mass of molecules, or any characteristic points that form a periodic set.
\smallskip

After data objects such as periodic crystals are introduced, the most important question of data science is \emph{Same or different?} \cite{sacchi2020same,coles2014same}, which appeared even in titles of crystallographic papers.
An objective answer to this question requires us to define what crystals can be called \emph{the same} by choosing a suitable equivalence relation.  
\medskip

\begin{dfn}[equivalence relation, equivalence class] 
\label{dfn:equivalence}
\textbf{(a)}
For any given objects, an \emph{equivalence} is a binary relation (denoted by $S\sim Q$) satisfying three axioms: 
\smallskip

\noindent
(1) \emph{reflexivity:} any object $S\sim S$;
\smallskip

\noindent
(2) \emph{symmetry:} if $S\sim Q$ then $Q\sim S$;
\smallskip

\noindent
(3) \emph{transitivity:} if $S\sim Q$ and $Q\sim T$ then $S\sim T$.
\medskip

\noindent
\textbf{(b)}
The \emph{equivalence class} $[S]$ of an object $S$ consists of all objects $Q$ equivalent to $S$.
\medskip

\noindent
\textbf{(c)}
One equivalence $S\sim_1 Q$ is called \emph{weaker} or \emph{coarser} than another $S\sim_2 Q$ (or $S\sim_2 Q$ is \emph{stronger} or \emph{finer} than $S\sim_1 Q$) if $S\sim_2 Q$ always implies that $S\sim_1 Q$.
\end{dfn}
\medskip

Any equivalence relation guarantees a well-defined classification by splitting all objects into disjoint equivalence classes. 
Indeed, if two classes $[S]$ and $[T]$ share a common object $Q$, then $[S]=[T]$ by
the transitivity axiom.
If an equivalence is not defined \emph{a priori} or axioms fail, all subsequent claims are hard to objectively justify.   
\smallskip

%For finite sets, one simple equivalence is by the size: $S\sim Q$ if $S,Q$ have equal sizes.
%Then each equivalence class $[S]$ contains all sets that consist of the same number (say, $m$) of points.
%Under this equivalence, the \emph{moduli space} of all finite non-empty sets is the collection of all positive numbers $m>0$.
For periodic crystals, one simple equivalence is by chemical composition: $S\sim Q$ if $S,Q$ have the same ratios of chemical elements.
Under this equivalence, all carbon allotropes such as diamond and graphite with vastly different properties fall into the same equivalence class.
To distinguish all polymorphs, % that can have different properties, 
we need a stronger (finer) equivalence that splits all periodic crystals into a wider variety of smaller classes.
\smallskip

On the other hand, a traditional CIF specifying a unit cell and a motif of atoms defines a too strict and impractical equivalence because one can shift all atoms of the motif by the same short vector even within a fixed unit cell and get a very different CIF of the same structure.
Hence a CIF is similar to a single snapshot (a 3D photo) but the same object can have infinitely many photos in different coordinate systems.
\smallskip

Crystallography tried to overcome this ambiguity by reducing a cell to a canonical form, e.g. Niggli cell \cite{niggli1928krystallographische}, and then by putting atoms at the most symmetric (Wyckoff) positions inside the fixed cell.
However, all reduced cells discontinuously change under almost any noise, which was reported experimentally in 1965 \cite[p.~80]{lawton1965reduced} and theoretically proved in 2022 \cite[Theorem~15]{widdowson2022average} even for lattices with empty motifs and any potential reduced cell.
Though this discontinuity was practically resolved in 2022 \cite[Theorems~4.3-4.4]{widdowson2022resolving}, many materials databases still accumulate near-duplicates. 
\smallskip
  
A practical mid-point compromise between coarse equivalences (by chemical composition or another property) and too strict coincidences (by cell-based CIFs or other coordinate-dependent representations) is a \emph{rigid motion} (transformation) justified by the fact that structures of solid crystalline materials are determined in a rigid form.  
%Experimental studies often rely on finite datasets whose entries are split into subjective classes without an objective equivalence. 
\medskip

\begin{exa}[rigid motion and isometry]
\label{exa:isometry}
%\textbf{(a)}
A \emph{rigid motion} is a composition of translations and rotations in $\R^n$.
An \emph{isometry} is a distance-preserving map $\R^n\to\R^n$.
A \emph{dilation} is a composition of isometry and uniform scaling in $\R^n$.
\end{exa}
\smallskip

In $\R^n$, any isometry is either a rigid motion or a composition of a mirror reflection with a suitable rigid motion and is also called a \emph{congruence} in Euclidean geometry, though an \emph{isometry} makes sense in more general metric spaces with non-Euclidean distances.
We use the term a \emph{rigid motion} rather than a rigid transformation because any such map $f_1:\R^n\to\R^n$ can be included into a continuous family (motion) of transformations $f_t:\R^n\to\R^n$, $t\in[0,1]$, connecting a given $f_1$ with the identify map $f_0=\mathrm{id}$ that fixes every point of $\R^n$.
Moreover, any isometry can also be called a rigid transformation in the sense that all inter-point distances are preserved.
The longer alternative name for a rigid motion is an \emph{orientation-preserving} isometry.
\smallskip

Under rigid motion, moduli spaces of crystals become naturally continuous because atomic coordinates have real continuous values, not restricted to a discrete grid.
\smallskip

\begin{dfn}[periodic structures and crystal structures] 
\label{dfn:structures}
In any dimension $n\geq 1$, a (rigid) \emph{periodic structure} is a class of periodic sets $S\subset\R^n$ of unlabelled points under rigid motion in $\R^n$.
In dimension $n=3$, if every point of a periodic set $S\subset\R^3$ has an atomic type, the class of $S$ under rigid motion is called a \emph{crystal structure}. 
\end{dfn}
\medskip

Under almost any noise or atomic vibration \cite{feynman2011lectures}, the class of a periodic crystal under rigid motion becomes slightly different.
Similarly, almost any movement of a physical object on the surface of Earth changes its geographic coordinates (latitude and longitude).
Crystal structures also live in continuous spaces defined below.     
\medskip

\begin{dfn}[crystal spaces]
\label{dfn:spaces}
The \emph{moduli space} under a given equivalence is the set of all equivalence classes.
For periodic sets of all atomic centers in a crystal, 
\smallskip

\noindent
the moduli space under rigid motion is called the \emph{Crystal Rigid Space ($\CRIS$)},
\smallskip

\noindent
the moduli space under isometry is called the \emph{Crystal Isometry Space ($\CIMS$)},
\smallskip

\noindent
the moduli space under dilation is called the \emph{Crystal Dilation Space ($\CRDS$)}.
\end{dfn}
\medskip

Fig.~\ref{fig:spaces-invariants}~(left) visualises spaces of crystals starting from finite collections of space-group types (230 in dimension 3) and chemical compositions, which form a much larger but still a finite and discrete set because ratios of (integer numbers of) chemical elements cannot change continuously while atomic coordinates certainly can. 
The space of rigid classes ($\CRIS$) is the largest in practice because there is no sense in excluding rotations or translations from an equivalence of crystals.
The weaker equivalence of isometries, which includes mirror reflections, defines a smaller (roughly speaking, twice smaller) space of isometry classes ($\CIMS$) by putting any mirror images into the same class under isometry.
The even weaker equivalence of dilation puts all uniformly scaled periodic sets, e.g. all cubic lattices, into the same class in the even smaller space $\CRDS$.
Any non-uniform scaling produces different classes under dilation.
\smallskip

In Definition~\ref{dfn:spaces} all points (zero-sized atomic centres) are considered unlabelled (indistinguishable).
If we add atomic types to points, we lose the ability to compare crystals with different chemical compositions.
Though it might seem that all chemistry is lost without atomic types, Richard Feynman's table in Figure 1-7 of his first lecture \cite{feynman2011lectures} was the Eureka moment for the last author in May 2021 to realise that precise enough geometry of unlabelled atomic centres should suffice to uniquely determine all chemical elements under fixed ambient conditions, such as temperature and pressure.  
\smallskip

Our physical intuition says that we cannot change atomic types at fixed geometric positions because interatomic interactions will shift atomic centres at least slightly (with two or three decimal places in Angstroms), which should be detectable by modern instruments. 
To rigorously justify that geometry can determine chemistry, we need reliable tools to practically distinguish all periodic point sets, not only known crystals. %chemically different
\smallskip

The moduli spaces in Definition~\ref{dfn:spaces} are not only continuous but are also infinite-dimensional, which makes distinguishing (separating) all periodic point sets a notoriously hard problem.  
If we fix (an upper bound of) a number $m$ of atoms in motifs of crystals, the resulting crystals spaces become finite-dimensional.
\smallskip

In $\R^3$, the spaces $\CRIS(\R^3;m)$ and $\CIMS(\R^3;m)$ have the dimension $3m$ due to $3$ coordinates for each of $m$ atoms and 6 parameters of a unit cell, which are neutralised by 6 parameters of rigid motion and isometry.
In practice, noise and atomic vibrations do not care about a fixed number $m$ of atoms and can easily scale up a primitive cell by any integer factor.
Hence, every periodic crystal with $m$ atoms in a unit cell is infinitesimally close to all (infinitely many) subspaces of crystals whose number of atoms in a unit cell is any integer multiple of $m$, see Fig.~\ref{fig:basis_ambiguity}~(bottom). 

%pro------------------------------------
\subsection{Mathematical definitions of invariants and metrics}
\label{sub:invariants+metrics}

The full problem statement needs a few concepts that are important in all sciences, e.g. Definition~\ref{dfn:invariants} helps rigorously answer the question \emph{Same or different?} 
For simplicity, we fix the main equivalence as rigid motion in $\R^n$, but most concepts and the subsequent geo-mapping problem make sense for other equivalences and objects. % beyond periodic crystals.
\medskip

\begin{dfn}[invariants and equivariants] 
\label{dfn:invariants}
\textbf{(a)}
An invariant $I$ is a function that takes as input any periodic point set $S\subset\R^n$ and outputs a value $I(S)$ that remains the same under any rigid motion $f$ in $\R^n$, i.e. $I(f(S))=I(S)$. 
Alternatively, if any periodic point sets $S,Q$ can be exactly matched by rigid motion (denoted as $S\cong Q$), then $I(S)=I(Q)$.
In other words, $I$ is a descriptor with \emph{no false negatives} that are pairs of different representations of rigidly equivalent sets: $S\cong Q$ with $I(S)\neq I(Q)$.
\medskip

\noindent
\textbf{(b)}
An equivariant $E$ is a function that takes as input any periodic point set $S\subset\R^n$ and outputs a value $E(S)$ that controllably changes under any rigid motion $f$ in $\R^n$, i.e. $E(f(S))=T_f(E(S))$, where a transformation $T_f$ depends on $f$ but not on $S$. 
\end{dfn}
\medskip

If $S$ is a periodic crystal, Definition~\ref{dfn:invariants} considers only rigid motions that also respect atomic types of $S$.
Our physical intuition says that this restriction is not practically needed, because replacing atomic types should change relative geometric positions of atomic centers at least slightly. % except small exceptions such as     
A rigorous justification of this conclusion requires complete invariants, which will be introduced in Definition~\ref{dfn:completeness} below.
\smallskip

The center of mass of all atoms in a unit cell $U$ (as a vector expressed in a basis of $U$) changes under any translation and hence is not invariant.
This center is also not equivariant in all dimensions  $n\geq 1$. 
%For example, 
The periodic sequence $S=\{0,1\}+2\Z$ has the centre $c=0.5(0+1)=0.5$ in the unit cell $[0,2)$.
The shift by $\ep>0$ to the right gives another representation $\{\ep,1+\ep\}+2\Z$ of $S$ (under translations) whose motif has the centre $c(\ep)=0.5(\ep+1+\ep)=0.5+\ep$ obtained by shifting $c=0.5$ by $\ep$ to the right.
The shift by $\ep>0$ to the left gives one more representation $\{2-\ep,1-\ep\}+2\Z$, where $-\ep$ was replaced with $2-\ep$ within the unit cell $[0,2)$. The new motif has the centre $c(-\ep)=0.5(2-\ep+1-\ep)=1.5-\ep$, which is not related to $c=0.5$ by any $\ep$-shift.
\smallskip

The basis vectors of a unit cell are equivariant but are not invariant and hence cannot distinguish any crystal structures under rigid motion.
In general, all linear combinations of equivariant vectors are also equivariant, so equivariants form a much wider collection than invariants, which explains that equivariant neural networks can be easily trained to fit any finite data.  
Invariants are much rarer in comparison with equivariants and hence do not allow easy overfitting.
\medskip

For example, the physical density of a periodic crystal $S$ is a scalar quantity that can be computed from any CIF of $S$.
Density satisfies the invariance, so a given crystal structure has one and only one density.
Hence, if crystals have different densities, their underlying structures are also different (cannot be exactly matched by rigid motion).
The important consequence of Definition~\ref{dfn:invariants}(a) is that any non-constant invariant distinguishes some (possibly, not all) crystals under rigid motion.
\medskip

\begin{dfn}[compete vs generically complete invariants] 
\label{dfn:completeness}
\textbf{(a)} 
An invariant $I$ is called \emph{complete} if any periodic point sets $S,Q\subset\R^n$ with equal invariant values $I(S)=I(Q)$ can be exactly matched by rigid motion in $\R^n$.
\medskip

\noindent
\textbf{(b)} 
An invariant $I$ is called \emph{generically complete} if the completeness property $I(S)=I(Q) \Rightarrow S\cong Q$ holds for almost all periodic point sets $S,Q\subset\R^n$ except sets that occupy a measure-zero subspace in the full space $\CRIS(\R^n)$ of rigid classes. 
\end{dfn}
\medskip

The completeness property means that our desired structure representation has a one-to-one mapping.
Given the value $I(S)$ of a complete invariant $I$ computed from a structure $S$, there is only one structure that corresponds to this $I(S)$.
Completeness is often a difficult property to satisfy.
We note in passing that, whilst it is a desirable property, non-complete continuous invariants, such as density, atomic pair distribution function (PDF), and even X-ray powder diffraction patterns (XRPD), are useful invariants!
Nonetheless, with that said, complete invariants are desirable to develop.
\medskip

\begin{dfn}[metrics vs pseudometrics] 
\label{dfn:metrics}
\textbf{(a)} 
A distance \emph{metric} $d$ on values of an invariant $I$ is a function satisfying the following metric axioms:
\smallskip

\noindent
(1) \emph{coincidence:} 
$d(a,b)=0$ if and only if $a=b$;
\smallskip

\noindent
(2) \emph{symmetry:} 
$d(a,b)=d(b,a)$; and
\smallskip

\noindent
(3) \emph{triangle inequality:} $d(a,b)+d(b,c)\geq d(a,c)$ for any values $a,b,c$.
\medskip

\noindent
\textbf{(b)} 
If condition (1) is replaced with the weaker one including one-directional implication
$(1')$ $d(a,a)=0$, then $d$ is called a \emph{pseudo-metric}.
\end{dfn}
\medskip

\begin{dfn}[$\ep$-perturbation and Lipschitz continuity] 
\label{dfn:continuity}
\textbf{(a)} 
For any $\ep>0$, periodic point sets $S,Q\subset\R^n$ are called $\ep$-perturbations of each other if there is a bijection $g:S\to Q$ such that $|g(p)-p|\leq\ep$ for any point $p\in S$.
\medskip

\noindent
\textbf{(b)}  
An invariant $I$ is called \emph{Lipschitz continuous} in a metric $d$ if there is a constant $\la$ such that, for any small enough $\ep>0$, if any periodic point sets $S,Q\subset\R^n$ are $\ep$-perturbations of each other, then $d(I(S),I(Q))\leq\la\ep$.  
\end{dfn}
\medskip

\subsection{A hierarchy of invariants: from the fastest to complete ones}
\label{sub:invariants}

We give simplified definitions of invariants and refer to more general versions in past papers.
Definition~\ref{dfn:PDD} introduces the main invariant $\PDD$ (Pointwise Distance Distribution) and its average vector $\AMD$ of Average Minimum Distances. 
\medskip

\begin{dfn}[$\AMD$ and $\PDD$ invariants]
\label{dfn:PDD}
Let $S\subset\R^n$ be any periodic set with a motif $M$ of $m$ points.
Fix an integer $k\geq 1$ and a point $p\in M$.
Let $d_1(p)\leq\dots\leq d_k(p)$ be the ordered list of Euclidean distances from $p$ to its $k$ nearest neighbours within the whole set $S$.
The unordered collection of these $m$ lists with equal weights $\frac{1}{m}$ for all points $p\in M$ is the \emph{Pointwise Distance Distribution} $\PDD(S;k)$.
The vector $\AMD(S;k)$ consists of the $k$ averages $\AMD_j(S)=\dfrac{1}{m}\sum\limits_{p\in M} d_j(p)$ for $j=1,\dots,k$.
\end{dfn}
\medskip

The PDD and AMD invariants can be defined for any discrete set $S=M$ of points in a metric space \cite[Definition~3.1]{widdowson2026pointwise} and are invariant under isometry in $\R^n$ \cite[Theorem~3.3]{widdowson2026pointwise}.
If any number (say, $l$) of distance lists in $\PDD(S;k)$ coincide, then can be collapsed into a single list with the \emph{weight} $\dfrac{l}{m}$.
Since all lists have the same length $k$, it is convenient to write the Pointwise Distance Distribution $\PDD(S;k)$ as a matrix with (up to $m$) unordered rows and $k$ ordered columns, plus an extra ($0$-th) column of weights.
The columns of $\PDD(S;k)$ are ordered by distance values, but rows are unordered because points of $S$ are unlabelled.
We also interpret $\PDD(S;k)$ as a distribution of weighted points $(d_1(p),\dots,d_k(p))$ in $\R^k$, one per point $p\in M$.
\medskip

One of the simple distance metrics on such distributions is Earth Mover's Distance ($\EMD$), which was previously applied to chemical compositions \cite{hargreaves2020earth}. 
Briefly, $\EMD$ optimally splits and matches rows of two $\PDD$s even if these distributions have different sizes and the same total (normalised) weight 1. 
For distributions of a maximum size $m$, the $\EMD$ can be computed exactly \cite{orlin1993faster} in time $O(m^3\log m)$ and approximately with a constant factor in time $O(m)$ \cite[section~3]{shirdhonkar2008approximate}.
If any points $p,q\in S$ are perturbed within their $\ep$-neighbourhoods in $\R^n$, then the Euclidean distance between them changes up to $2\ep$ due to the triangle inequality in Definition~\ref{dfn:metrics}(a).
Extending this basic fact, \cite[Theorem~4.3]{widdowson2022resolving} proved that the $\PDD$ is continuous in the $\EMD$ with Lipschitz constant $2$ for all periodic point sets and small enough perturbation $\ep>0$. 
\medskip
%\end{comment}

$\PDD(S;k)$ may look similar to the Pair Distribution Function (PDF), which is a distribution of all inter-atomic distances \cite{billinge2023atomic}.
This infinite discrete distribution is often normalised by frequency and further smoothed by Gaussian blurring to avoid discontinuities under scaling up and perturbing a unit cell in Fig.~\ref{fig:basis_ambiguity}~(top right).
Even if PDF is split by atomic types, the resulting invariant is weaker than the PDD, which is split by all (isometrically non-equivalent) points, hence depending only on geometry rather than on chemistry, see \cite[Example~3.3]{widdowson2022resolving} and the subsequent discussion of PDF.
\medskip 
 
The key advantage of PDD is its generic completeness  \cite[Theorem~4.4]{widdowson2022resolving} in the sense of Definition~\ref{dfn:completeness}(b).
Detailed proofs of all results about the PDD are in \cite{widdowson2026pointwise}, which also extended this invariant to more general $l$-periodic sets $S\subset\R^n$ with periodicity in $l$ directions within $\R^n$ for any $1\leq n$.
For $l=1$, complete and continuous invariants of 1-periodic high-dimensional sequences in $\R^n$ were designed in \cite{kurlin2025complete}.
\medskip 

Definition~\ref{dfn:PDDh} extends the $\PDD$ to a hierarchy of higher-order invariants $\PDDh$ that capture geometric information about relative positions of $h+1\geq 3$ atoms
\cite{widdowson2027higher}.
\medskip 

\begin{dfn}[higher -order $\PDDh(S;k)$] % and $\PDDall$]
\label{dfn:PDDh}
Let $S\subset\R^n$ be any periodic set with a motif $M$ of $m$ points.
Fix a point $p\in M$ and integers $h,k\geq 1$.
Consider any $h$ distinct points $p_1,\dots,p_h\in S\setminus\{p\}$ and the \emph{$h$-order average} $\dfrac{2}{h(h+1)}\sum\limits_{0\leq i<j\leq h}|p_i-p_j|$ of pairwise distances between the points $p=p_0,p_1,\dots,p_h$. 
Let $a(p;h,1)\leq\dots\leq a(p;h,k)$ be the list of $k$ smallest averages for the fixed point $p$ and variable points $p_1,\dots,p_h\in S\setminus\{p\}$.
The unordered collection of these lists with equal weights $\dfrac{1}{m}$ for all points $p\in M$ is called the \emph{$h$-order Pointwise Distance Distribution} $\PDD^{\{h\}}(S;k)$.
\end{dfn}
\medskip 

The invariant $\PDD^{\{2\}}(S;k)$ distinguished all  (infinitely many) known counterexamples to the completeness of the PDD through manual computations in \cite[Example~3.4]{widdowson2027higher}, which demonstrates its simplicity in addition to strength.
All invariants $\PDDh(S;k)$ are continuous with Lipschitz constant 2 \cite[Theorem~4.3]{widdowson2027higher} and computable in polynomial time, for the fixed $h,n$ \cite[Theorem~5.10]{widdowson2027higher}.
\medskip

\begin{dfn}[$\al$-clusters and isoset $I(S;\al)$]
\label{dfn:isoset}
For a point $p$ in a periodic point set $S\subset\R^n$ and any radius $\al\geq 0$, the \emph{$\al$-cluster} $C(S,p;\al)$ is the set of all vectors $\vec q - \vec p$ such that $q\in S$ and $|\vec q - \vec p|\leq\al$.
The \emph{isoset} $I(S;\al)$ is the unordered collection of all isometry classes $[C(S,p;\al)]$ of $\al$-clusters with equal weights for all points $p$ in a motif $M$ of $S$.
If we replace isometry with rigid motion, we get the \emph{oriented} isoset $I^o(S;\al)$.
\end{dfn}
\medskip

In comparison with $\PDD(S;k)$, a local cluster around a point $p$ keeps all vectors $\vec q - \vec p$ for neighbours $q$ without reducing them to distances.
Since each cluster $C(S,p;\al)$ is not invariant under rotations around $p$, the isoset $I(S;\al)$ is necessarily defined via isometry classes of clusters. 
In practice, this abstraction is resolved by computing an approximation to the Lipschitz continuous metric between isosets, which needs only representative clusters, not their invariants, see details in \cite[sections~4-5]{anosova2026recognition}.    

%B======================
\section{More experimental details on the GNoMEs}
\label{sec:GNoME_details}

Table~\ref{tab:GNoME_release_notes} shows the dates and release notes of the four GNoME versions from \cite{gnome2023data}.
These updates withdrew entries as well as added them, but no note is given of withdrawal counts.
The only quantitative characterisation of any update is ``significantly increasing the number of released crystals''. 
The overview still describes the dataset as ``$\sim$381,000 novel structures'', although the latest version has 554,054 entries. 
\smallskip

\begin{table}[h!]
\caption{Dates and public notes about the four GNoME versions, quoted from \cite{gnome2023data}.}
\label{tab:GNoME_release_notes}
\begin{tabular}{rp{100mm}}
dates & release notes of GNoME versions \\
\hline
29 November 2023 & ``Initial dataset release''\\
1 December 2023 & ``Re-introduce paper filters to remove un-physical energies; add 2 missing columns'' \\
21 August 2024 & ``Adjust threshold to 1meV/atom off the hull significantly increasing the number of released crystals, update to use `require\_bound' version of MP corrections, re-relax structure where VASP did not update lattice coordinates, provide convex hull entries from recomputations of MP / OQMD'' \\
21 November 2024 & ``Ensure consistency in formation energies, provide used elemental references + pseudopotential corrections'' 
\end{tabular}
\end{table}
 
Old GNoME versions are online but are omitted from the documented directory structure. 
The release notes give a directory tree for the public bucket gs://gdm\_materials\_discovery, which does  contain gnome\_data/legacy\_versions/, holding subdirectories 11\_29\_23/, 12\_1\_23/ and 8\_21\_24/ with the summary tables and, for the latter two, full structure archives. 
But this directory appears nowhere in the release notes. 
The earlier releases are therefore public but undiscoverable from the documentation, which is why comparisons of the kind reported here have not been made: reproducing them requires knowledge of a path that the documentation omits.
\smallskip

%B.1-----------------------------------
\subsection{Details of invariant-based comparisons of the GNoMEs}
\label{sub:GNoME_comparisons_details}

We ran all experiments on modest AMD Ryzen 5 5600X 6-core, 32GB RAM.
\medskip

While geometrically matching entries up to thresholds $\tau\leq 10^{-5}\angstrom$, all groups of matched entries in GNoME 1\&2 are \emph{cliques}, i.e., all pairs of members of a group are within $\tau$. %, so group membership is unambiguous.
At $\tau=10^{-2}$, only 9,428 groups (82,224 entries) in GNoMEs 1\&2 are not cliques and may have the maximum distance among members larger than $\tau$. %inflate some results below which can be considered upper bounds.
\smallskip

\emph{Screening}. Pairs were screened in three blocks (those in GNoME1 only, in GNoME2 only, and across the two), initially by RMS on $\ADA[100]$. The largest threshold $\tau=10^{-2}$ returned 134,686,519 pairs, with the GNoME2 only block (added entries) being the densest. Survivors were then further compared in the sequence $\PDA[100] \to \PDAsq[100] \to \text{isosets}$, each stage removing pairs whose distance exceeded $\tau$. 
\smallskip

Exactly 1,453 screened pairs were trivial (known identical CIFs) and were passed through without comparison. Of the 134,685,066 pairs compared by $\PDA[100]$, 8,010,884 (5.9\%) survived to $\PDAsq[100]$ and 1,344,998 (1.0\% of all screened) to the isoset stage, of which 242,010 (18.0\%) were within a threshold $\tau$ at every stage. With the trivial pairs added back, 243,463 pairs had all distances within $\tau=10^{-2}$.
\smallskip

At $\tau = 10^{-10}$, GNoME1 has 3,248 entries (0.84\%) in 1,567 geometric duplicate groups.
These groups were previously found by comparing CIFs as texts in \cite[Table~2]{anosova2024importance}, which amounts to 1,681 redundant copies, a redundancy rate of 0.437\%. 
\smallskip

\begin{table}[h!]
  \centering
  \caption{Pair counts at each stage of the comparison process for the threshold $\tau = 10^{-2}$. Each invariant column gives the number of pairs compared at that stage. Trivial pairs (known identical CIFs) were passed through costing nothing. The \emph{screened} column contains the number of pairs which passed the initial comparison by $\ADA[100]$.}
  \label{tab:cascade-blocks}
  \setlength{\tabcolsep}{4pt}
  \small
  \begin{tabular}{lrr|rrr|r}
    & & & \multicolumn{3}{c|}{pairs compared by invariants} & geometric \\
    block        & screened    & trivial & $\PDA$      & $\PDAsq$  & isosets   & matches \\ \hline
    GNoME1 only & 30,732,748  & 1,224   & 30,731,524  & 2,058,581 & 420,611   & 84,014  \\
    GNoME2 only & 40,861,239  & 0       & 40,861,239  & 2,594,823 & 420,282   & 76,475  \\
    Both         & 63,092,532  & 229     & 63,092,303  & 3,357,480 & 504,105   & 82,974  \\ \hline
    total        & 134,686,519 & 1,453   & 134,685,066 & 8,010,884 & 1,344,998 & 243,463 \\
  \end{tabular}
\end{table}

GNoME2 contains only 591 such entries in 273 groups, 318 redundant copies and redundancy rate 0.057\%, a more than seven-fold drop in rate despite GNoME2 being 44\% larger, and the gap survives every threshold up to $10^{-3}$ in Tables~\ref{tab:GNoME1_duplication} and~\ref{tab:GNoME2_duplication}. 
\smallskip

For geochemical matches, GNoME1 holds 1,133 (at $\tau = 10^{-10}\angstrom$) and 4,172 (at $\tau = 10^{-2}\angstrom$) geochemical groups, giving 2,311 and 8,622 entries in total, whereas GNoME2 contains no geochemical matches up to $\tau=0.01\angstrom$.
\medskip

\begin{table}[h!]
  \centering
  \caption{Computational times.  Every pair is screened once at the largest threshold $\tau=10^{-2}\angstrom$.
 Results for smaller thresholds can then be derived in trivial time. The experiment ran on 8 worker processes,
  so times are in CPU hours summed over workers and are quoted as
  percentages of that sum. The total wall-clock time taken to complete comparisons was 16 hours, 11 minutes and 54 seconds.}
  \label{tab:times}

  \begin{tabular}{lrrr}
    Stage & comparisons & CPU time (hours:min:sec) & \% of total \\
    \hline
    EMD on $\PDA[100]$                & 134,685,066 & 30:19:12 & 26.6  \\
    EMD on $\PDAsq[100]$              & 8,010,884   & 30:51:52 & 27.1  \\
    EMD on $\isoset[5]$       & 1,344,998   & 52:52:50 & 46.4  \\
    \hline
    total over 8 workers     &             & 114:03:45 & 100.0 \\
  \end{tabular}
\end{table}

\begin{figure}[h!]
    \centering
    \includegraphics[width=\linewidth]{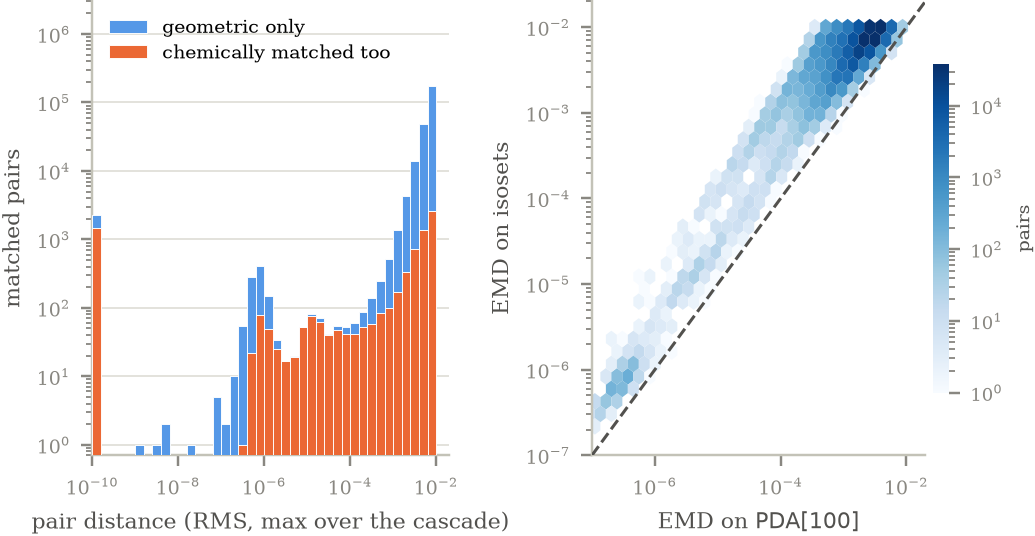}
    \caption{\textbf{Left}: distribution of distances between all pairs compared in GNoME 1\&2, taken as the maximum distance over four invariants. 
\textbf{Right}: $\EMD$ on $\PDA[100]$ vs $\EMD$ on isoset[5].}
    \label{fig:GNoMEs_EMD_PDA100-vs-isosets5}
\end{figure}

\begin{figure}[h!]
    \centering
    \includegraphics[width=\linewidth]{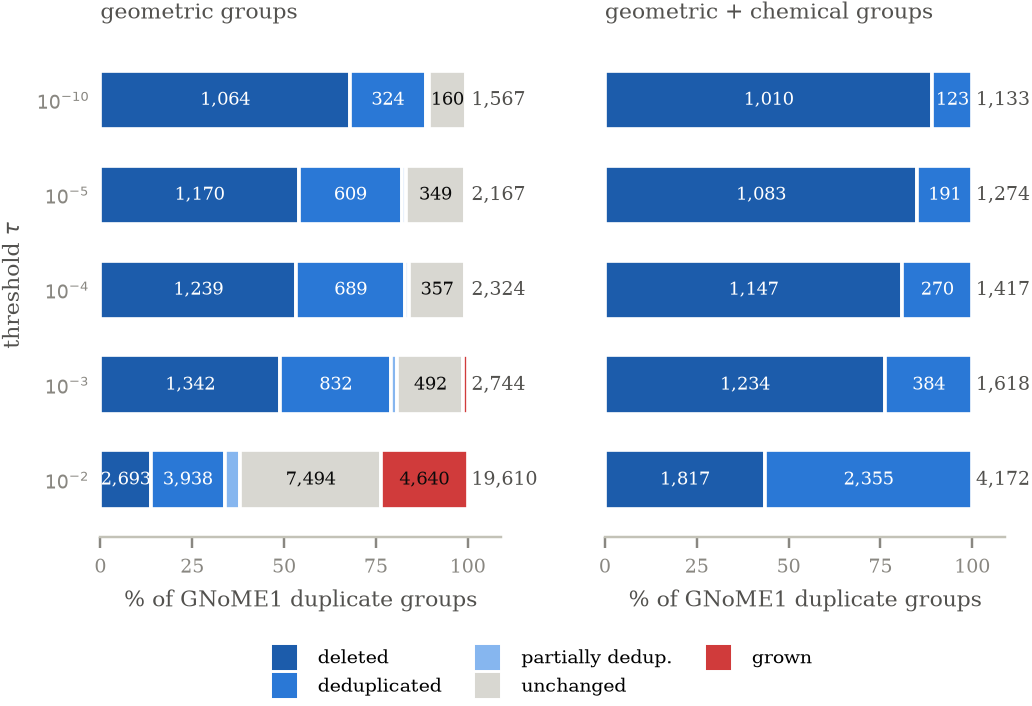}
    \caption{Split of the fates of the various groups found in GNoME1, as described in Table~\ref{tab:GNoME1_groups_fates}.}
    \label{fig:GNoME1_groups_fates}
\end{figure}

\begin{figure}[h!]
    \centering
    \includegraphics[width=\linewidth]{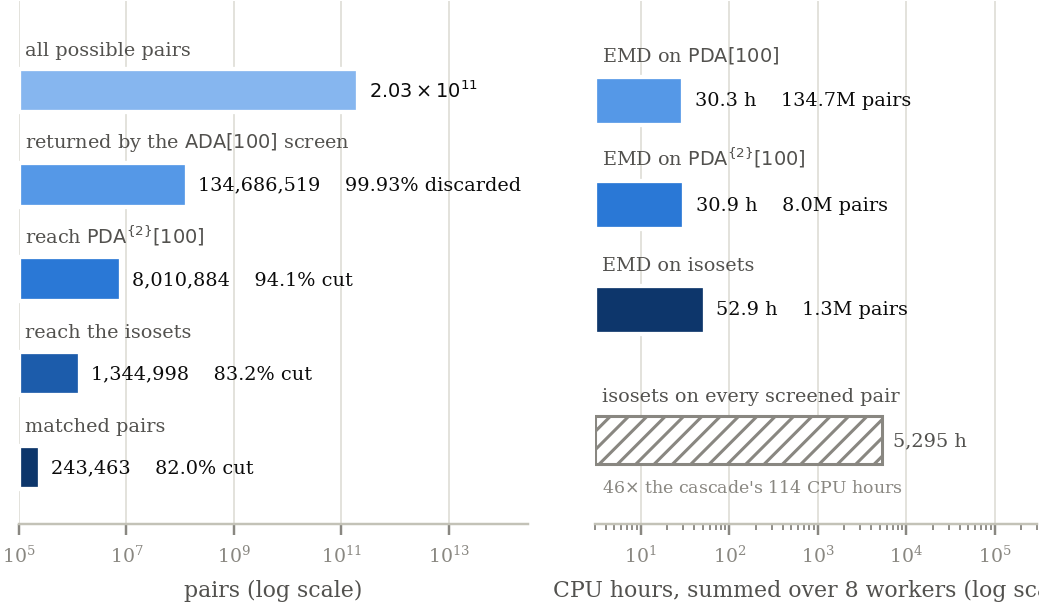}
    \caption{
    \textbf{Left}: Number of pairs and percentage cut at each stage of comparisons. 
    \textbf{Right}: Time of each stage (Table~\ref{tab:times}), and estimated time to compare all pairs passing the $\ADA[100]$ screen by isosets.}
    \label{fig:pairs_and_times}
\end{figure}

For GNoMEs 1\&2, duplication predicts removal very strongly: 95.3\% of the GNoME1 entries (2,202 of 2,311) with a geochemical duplicate were dropped at $\tau=10^{-10}\angstrom$, and 73.9\% (6,373 of 8,622) at $10^{-2}$, against a background removal rate of 21.7\% (83,525 of 384,938), but the converse is not true. Only 626 of the 83,525 removed entries (0.75\%) had a geometric match that survived into GNoME2 at $\tau = 10^{-10}$, and only 367 (0.44\%) had a geochemical one; even at $\tau = 10^{-2}$, the figures are 11,547 (13.8\%) and 5,169 (6.2\%). 
The bulk of the 83,525 removals were unique structures.

\begin{table}[htbp]
  \centering
  \caption{Curation outcome against duplication status. Each entry is grouped by its multiplicity in its own source release
  (GNoME2 for added entries, GNoME1 for removed and retained ones). Removal is strongly concentrated on the duplicated entries: at the threshold $\tau = 10^{-10}$, 95.3\% of the GNoME1 entries with a geochemical duplicate were dropped, against a background
  removal rate of 21.7\%.}
\label{tab:GNoME1_entries_fates}
  \begin{tabular}{llrrrr}
    $\tau, \angstrom$ & status & unique structures & geometric duplicates & geochemical duplicates & total \\ \hline
    \multirow{3}{*}{$10^{-10}$}
      & added    & 252,473 & 168    & 0     & 252,641 \\
      & removed  & 80,949  & 374    & 2,202 & 83,525  \\
      & retained & 300,741 & 563    & 109   & 301,413 \\ \hline
    \multirow{3}{*}{$10^{-5}$}
      & added    & 252,308 & 333    & 0     & 252,641 \\
      & removed  & 80,414  & 690    & 2,421 & 83,525  \\
      & retained & 300,028 & 1,208  & 177   & 301,413 \\ \hline
    \multirow{3}{*}{$10^{-4}$}
      & added    & 252,300 & 341    & 0     & 252,641 \\
      & removed  & 80,190  & 703    & 2,632 & 83,525  \\
      & retained & 299,932 & 1,226  & 255   & 301,413 \\ \hline
    \multirow{3}{*}{$10^{-3}$}
      & added    & 251,968 & 673    & 0     & 252,641 \\
      & removed  & 79,740  & 847    & 2,938 & 83,525  \\
      & retained & 299,424 & 1,621  & 368   & 301,413 \\ \hline
    \multirow{3}{*}{$10^{-2}$}
      & added    & 197,073 & 55,568 & 0     & 252,641 \\
      & removed  & 67,639  & 9,513  & 6,373 & 83,525  \\
      & retained & 249,811 & 49,353 & 2,249 & 301,413 \\
  \end{tabular}
\end{table}

\begin{table}[htbp]
  \centering
  \caption{Entry turnover across the four GNoME releases, computed by matching \texttt{MaterialId} between the \texttt{stable\_materials\_summary.csv} of each version. \emph{Withdrawn} counts identifiers present in the earlier release and absent from the later one. The releases are not nested: 10,201 entries that were withdrawn on 8/21/24 were restored on 11/21/24.}
  \label{tab:GNoME_changes}
  \begin{tabular}{lrrrr}
& &  \multicolumn{3}{c}{change from the previous version} \\
    date & entries & withdrawn entries & added entries & net change \\
    \hline
    11/29/23 & 384,938 & --- & --- & --- \\
    12/1/23  & 384,871 & 67 & 0 & $-67$ \\
    8/21/24  & 529,803 & 92,095 & 237,027 & $+144{,}932$ \\
    11/21/24 & 554,054 & 15,615 & 39,866 & $+24{,}251$ \\
    \hline
    \multicolumn{2}{l}{11/29/23 $\rightarrow$ 11/21/24} & 83,525 & 252,641 & $+169{,}116$ \\
  \end{tabular}
\end{table}

%B.2--------------------------
\subsection{Details of comparing the GNoMEs against ICSD and MP}
\label{sub:GNoMEs-vs-ICSD-MP}

This section describes findings from comparisons of both GNoME releases against the Inorganic Crystal Structure Database (ICSD) and Materials Project (MP). Steps taken in carrying out the comparisons are laid out in Section~\ref{subsection:methods}.
\medskip

We took ICSD and MP version \cite{MP2026} with the parameter “include gnome=False”, and ignored all entries deposited from March 2021, when GNoME started training.
\smallskip

\noindent
- v2024.11.14 (released 12 Dec 2024): "The current set of 2,047 GNoME-originated materials has been deprecated in preparation for a release of about 120,000 GNoME materials"
\smallskip

\noindent
- v2024.12.18 (released 20 Dec 2024): "Added 15,483 GNoME-originated materials calculated using r2SCAN ... A further release of an additional ~100k GNoME materials is in preparation."
\smallskip

\noindent
- v2025.04.10 (released 18 April 2025): "Added 30,00 GNoME-originated materials calculated using r2SCAN"
\smallskip

\noindent
- v2026.04.13 (released 8th June 2026): "Added 74,052 new GNoME materials computed at the r2SCAN level of theory. This brings the total to 117k GNoME materials."
\medskip

\emph{GNoME's geometry is largely the geometry of known crystals}. 
Table~\ref{tab:GNoMEs-vs-ICSD_MP_distances} gives the distribution of the distance from each entry to its nearest reference structure by geometry alone. The median GNoME2 entry sits $0.034\angstrom$ from some ICSD structure and $0.03\angstrom$ from some MP structure. 
This might be explained by GNoME's generative approach, which takes existing crystals and relabels chemical elements with similar ones.
\medskip

\emph{Chemical matches are rare, and the population that could produce one is small}. 
Only $29{,}167$ GNoME2 entries ($5.3\%$) have any ICSD entry with the same element set, and only $5{,}201$ ($0.9\%$) have one with the same composition.
For the Materials Project (MP), the corresponding figures are $33{,}566$ and $3{,}887$.
In GNoME2, only $824$ entries match an ICSD crystal within $0.05\angstrom$ by geometry and chemistry together, and only $1$ entry matches an MP crystal, totalling $0.15\%$ of GNoME2 (Table~\ref{tab:ref-counts-g2}). 
The corresponding GNoME1 figures are $1{,}315$ and $155$, totalling $0.37\%$ (Table~\ref{tab:ref-counts-g1}).
\medskip

\emph{Duplication of known crystals dropped sharply between releases}. 
Of the $559$ GNoME1 entries that were geochemical matches to a known \emph{ordered} crystal within $0.05$, $92.7\%$ were withdrawn in GNoME2 (against a background removal rate of $21.7\%$); of the $155$ matching a crystal in the MP $99.4\%$ were withdrawn, leaving one (Table~\ref{tab:ref-fates}).
\medskip

\emph{Removal rates are very different for ordered and disordered entries}. Matches to a known \emph{ordered} crystal were removed at a rate of $92.7\%$, but matches to a \emph{disordered} one at only $40.4\%$, against the $21.7\%$ background removal rate. This difference leads us to believe that a filter was applied to remove known ICSD crystals from GNoME1, but that it was unable to identify entries matching a disordered ICSD crystal. 
\smallskip

GNoME2 still contains $774$ such entries against GNoME1's $879$, a fall of $12\%$ where matches to ordered crystals fell by $89\%$.

%\subsection{Tables (comparisons vs ICSD, MP)}

\begin{table}[htbp]
  \centering
  \caption{GNoME1 against ICSD and MP references. Each entry is assigned the first category that applies at a threshold $\tau$: \emph{ordered}, a chemically constrained (geochemical) match to an ordered reference crystal; \emph{disordered} (details in subsection~\ref{subsection:methods}), and \emph{only geometric}, a match by $d_{\text{geom}}$ alone.}
  \label{tab:ref-counts-g1}
  \setlength{\tabcolsep}{6pt}
  \begin{tabular}{lrrrrr}
%    &         &            & any      & geometry & \\
    threshold $\tau$ & ordered & disordered & any chemical & only geometric & none \\
    \hline
    \multicolumn{6}{l}{ICSD} \\
    $0.005$ & 23  & 35    & 58    & 4,213  & 380,667 \\
    $0.01$  & 60  & 90    & 150   & 12,915 & 371,873 \\
    $0.02$  & 162 & 241   & 403   & 60,317 & 324,218 \\
    $0.03$  & 288 & 475   & 763   & 136,240 & 247,935 \\
%    $0.05$  & 436 & 879   & 1,315 & ---    & ---     \\
%    $0.1$   & 597 & 1,406 & 2,003 & ---    & ---     \\
    \hline
    \multicolumn{6}{l}{MP (no disordered structures)} \\
    $0.005$ & 30  & 0 & 30  & 5,560  & 379,348 \\
    $0.01$  & 57  & 0 & 57  & 19,818 & 365,063 \\
    $0.02$  & 89  & 0 & 89  & 82,513 & 302,336 \\
	$0.03$  & 112 & 0 & 112 & 165,815 & 219,011 \\
%    $0.05$  & 155 & 0 & 155 & ---    & ---     \\
%    $0.1$   & 216 & 0 & 216 & ---    & ---     \\
  \end{tabular}
\end{table}

\begin{table}[htbp]
  \centering
  \caption{GNoME2 against ICSD and MP; columns as in Table~\ref{tab:ref-counts-g1}. Matches to an ordered reference crystal have mostly disappeared, while matches to a disordered one have not.}
  \label{tab:ref-counts-g2}
  \setlength{\tabcolsep}{6pt}
  \begin{tabular}{lrrrrr}
%    &         &            & any      & geometry & \\
    threshold $\tau$ & ordered & disordered & any chemical & only geometric & none \\
    \hline
    \multicolumn{6}{l}{ICSD} \\
    $0.005$ & 9  & 27    & 36    & 4,289   & 549,729 \\
    $0.01$  & 20 & 69    & 89    & 17,325  & 536,640 \\
    $0.02$  & 29 & 201   & 230   & 93,382  & 460,442 \\
    $0.03$  & 36 & 413   & 449   & 214,404  & 339,201   \\
%   $0.05$  & 50 & 774   & 824   & ---     & ---     \\
%    $0.1$   & 67 & 1,220 & 1,287 & ---     & ---     \\
    \hline
    \multicolumn{6}{l}{MP (no disordered structures)} \\
    $0.005$ & 0 & 0 & 0 & 6,552   & 547,502 \\
    $0.01$  & 0 & 0 & 0 & 30,689  & 523,365 \\
    $0.02$  & 0 & 0 & 0 & 134,431 & 419,623 \\
    $0.03$  & 1 & 0 & 1 & 264,765 & 289,288     \\
%    $0.05$  & 1 & 0 & 1 & ---     & ---     \\
%    $0.1$   & 1 & 0 & 1 & ---     & ---     \\
  \end{tabular}
\end{table}

\begin{table}[htbp]
  \centering
  \caption{What became of the GNoME1 entries that matched a known crystal, at $\tau = 0.05$ and against either reference set. Matches to a known ordered crystal were removed at four times the background removal rate of $21.7\%$, but only $40.4\%$ of matches to disordered crystals were removed and 226 were added.}
  \label{tab:ref-fates}
  \setlength{\tabcolsep}{6pt}
  \begin{tabular}{lrrr}
    fate of crystals & ordered & disordered & any chemical \\
    \hline
    retained (301,413)  & 41  & 508 & 549 \\
    removed (83,525)    & 518 & 345 & 863 \\
    added (252,641)     & 10  & 266 & 276 \\
  \end{tabular}

  \medskip

  \begin{tabular}{lrr}
    GNoME1 category at the threshold $\tau = 0.05$ & entries & removed (\%) \\
    \hline
    ordered                        & 559     & 92.7 \\
    disordered                     & 853     & 40.4 \\
    any chemical match in ICSD     & 1,315   & 58.3 \\
    any chemical match in MP       & 155     & 99.4 \\
    no chemical match              & 383,526 & 21.6 \\
  \end{tabular}
\end{table}

\begin{table}[htbp]
  \centering
  \caption{The closest matches to known crystals surviving in GNoME2, at $\tau = 0.05\angstrom$. An asterisk * marks an entry added in GNoME2. 
Upper panel: geochemical matches to a periodic reference crystal, $d_{\text{chem}}$. Lower panel: matches to a disordered ICSD crystal; the twelve closest are all ordered.}
  \label{tab:ref-examples}
  \setlength{\tabcolsep}{5pt}
  \begin{tabular}{lllr}
    GNoME id & composition & reference & distance, $\angstrom$ \\
    \hline
    \multicolumn{4}{l}{\emph{ordered matches} (50 in GNoME2, 18 of them to experimental ICSD)} \\
    f7ee908e60* & IrSnTa        & ICSD 689927 & 0.000009 \\
    c5f1363c48  & GaOSc$_3$     & ICSD 687606 & 0.000066 \\
    9993fb827a  & InOSc$_3$     & ICSD 687607 & 0.000154 \\
    1cd0016832  & AuNNi$_3$     & ICSD 680114 & 0.000296 \\
    9d280540c9* & NNi$_3$Pd     & ICSD 680121 & 0.000370 \\
    12b40befa0* & Co$_3$NPd     & ICSD 680008 & 0.000672 \\
    bba5df2ac3* & Co$_3$CuN     & ICSD 680017 & 0.000702 \\
    f4e72c3b19  & AuCo$_3$N     & ICSD 680006 & 0.000823 \\
    b39b607649  & H$_6$BaIrK    & ICSD 262197 & 0.003956 \\
    cd251c1afc  & Ca$_4$P$_2$Se & ICSD 686117 & 0.005293 \\
    8a2692b9bd  & CFe$_2$LuSi   & mp-1207076  & 0.022086 \\
    \hline
    \multicolumn{4}{l}{\emph{matches to a disordered ICSD crystal} (774 in GNoME2, 266 of them new)} \\
    d1448232be  & Os$_3$V$_{13}$      & ICSD 647852, Os$_{0.4}$V$_{1.6}$          & 0.000762 \\
    6d759f71af  & Al$_8$DyEr$_3$      & ICSD 291821, AlDy$_{0.2}$Er$_{0.3}$       & 0.001673 \\
    1d6f38bd7b* & AsSSm$_2$           & ICSD 611321, As$_{0.55}$S$_{0.45}$Sm      & 0.001980 \\
    c5b9a84fae  & Al$_4$Co$_3$Fe      & ICSD 606563, AlCo$_{0.7}$Fe$_{0.3}$       & 0.001992 \\
    4fef57a136  & CrFe$_{15}$         & ICSD 692954, Cr$_{0.04}$Fe$_{1.96}$       & 0.002034 \\
    36817ff8d9  & Al$_8$DyHo$_3$      & ICSD 291824, AlDy$_{0.2}$Ho$_{0.3}$       & 0.002052 \\
    8d0043ba5a  & Be$_5$Co$_3$        & ICSD 058694, Be$_{1.34}$Co$_{0.66}$       & 0.002201 \\
    8666ed5124  & Al$_8$Dy$_3$Er      & ICSD 291819, AlDy$_{0.4}$Er$_{0.1}$       & 0.002380 \\
    67e4bb6449  & Al$_8$Dy$_3$Ho      & ICSD 291823, AlDy$_{0.4}$Ho$_{0.1}$       & 0.002558 \\
    1c3630d10b  & Fe$_{15}$V          & ICSD 634026, Fe$_{1.92}$V$_{0.078}$       & 0.002579 \\
    6fbb16782c  & Nb$_5$Re$_3$        & ICSD 105210, Nb$_{1.2}$Re$_{0.8}$         & 0.003143 \\
    a4de687109* & Co$_{11}$O$_{16}$Zn & ICSD 073754, Co$_{0.605}$OZn$_{0.145}$    & 0.003206 \\
  \end{tabular}
\end{table}

\begin{figure}[h!]
\centering
\includegraphics[height=45mm]{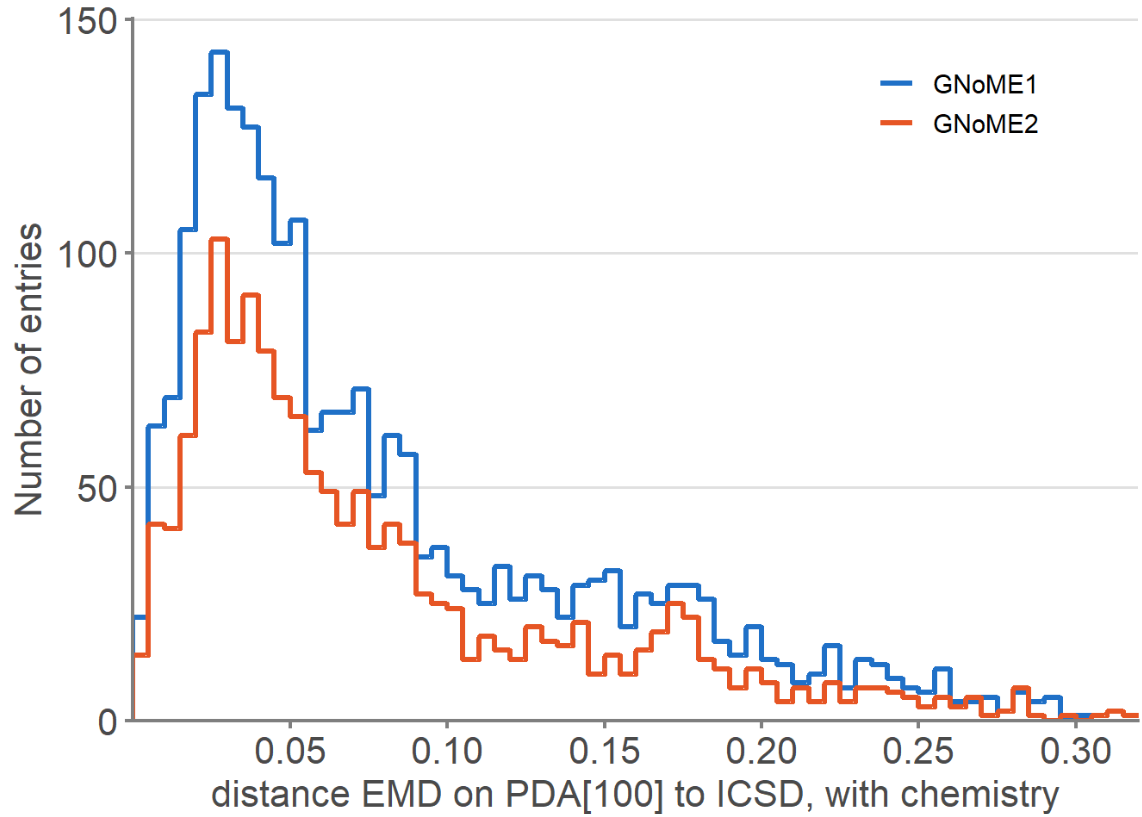}
\includegraphics[height=45mm]{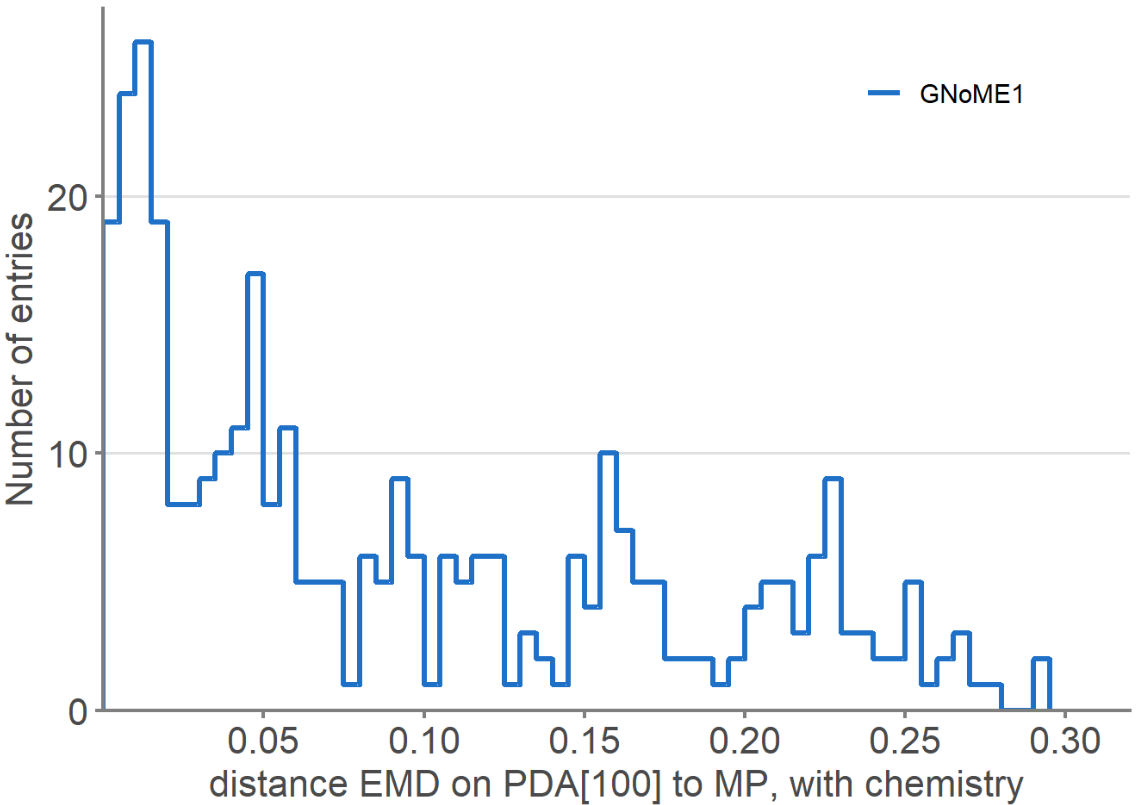}
\smallskip

\includegraphics[height=45mm]{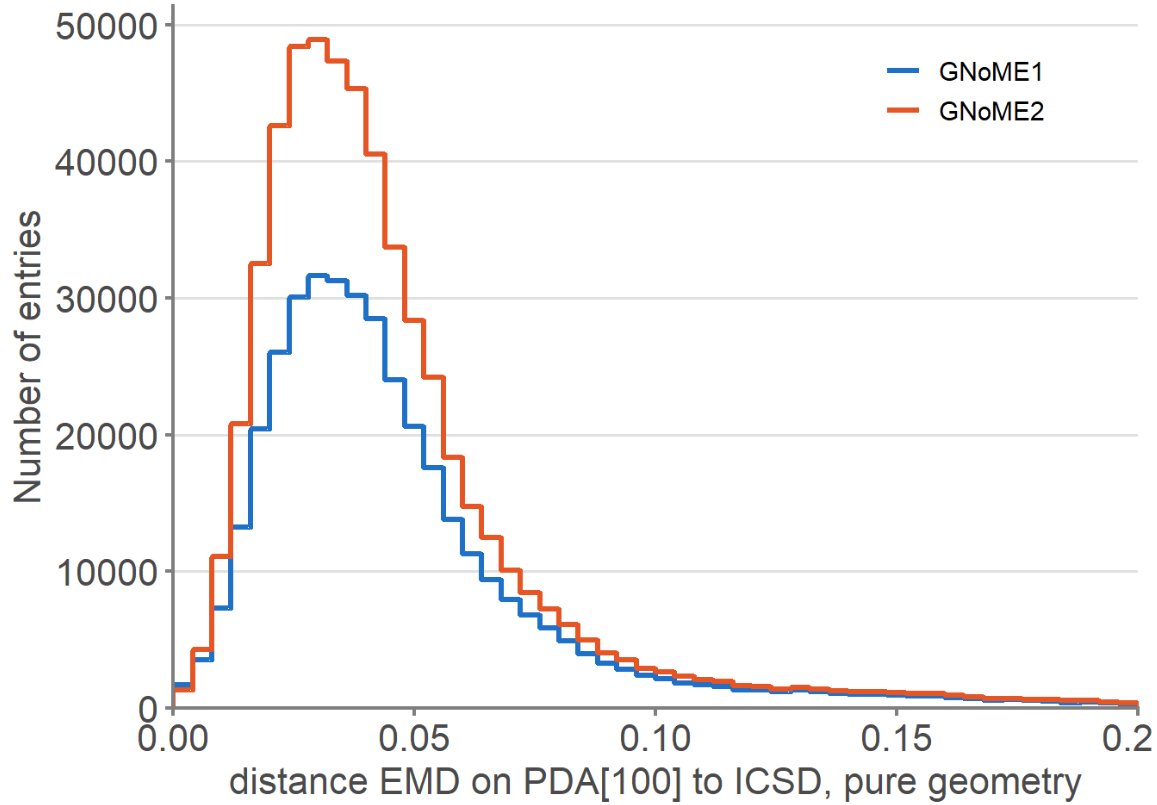}
\includegraphics[height=45mm]{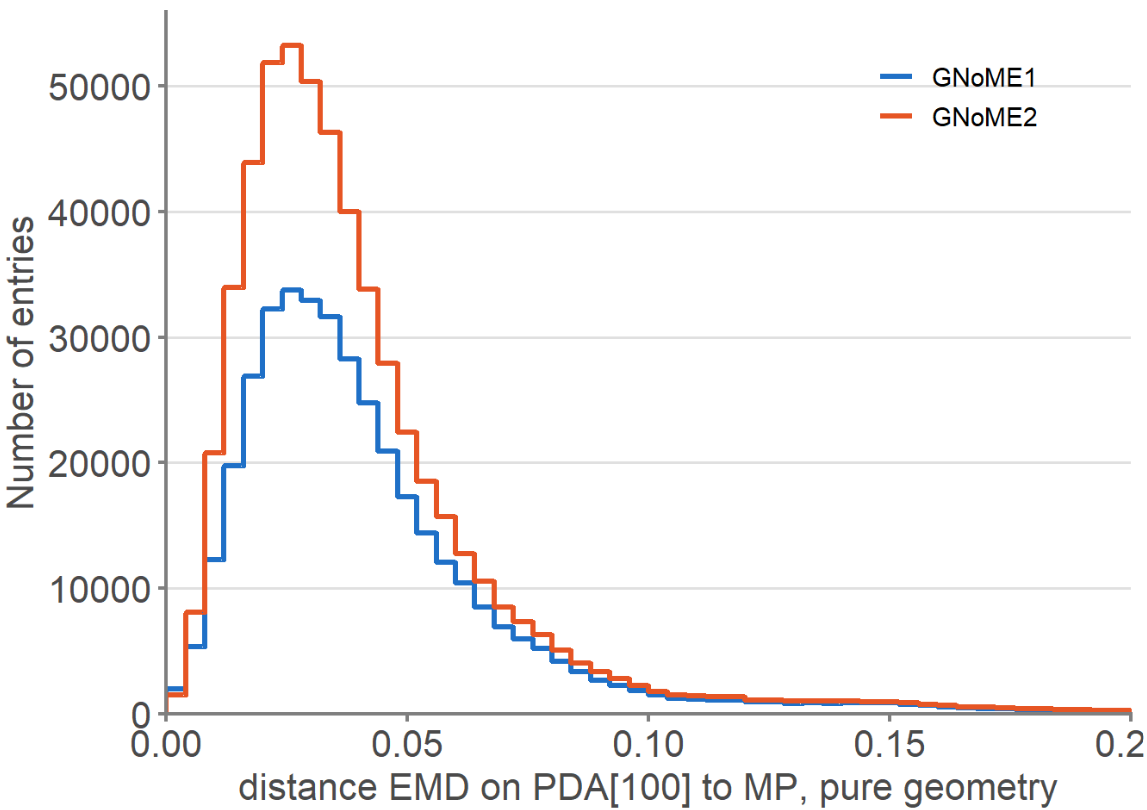}
\caption{
Histograms of distances based on $\EMD$ between invariants $\PDA[100]$, for predictions in GNoMEs 1\&2 to the nearest neighbours among known crystals in the ICSD and MP. 
\textbf{Left}: ICSD.
\textbf{Right}: MP.
\textbf{Top}: distance $d_{\text{chem}}$ matches chemistry, see Table~\ref{tab:comparison_distances}. 
The curve on the right is for GNoME1 only, because GNoME2 has no chemical matches in the MP.
\textbf{Bottom}: distance $d_{\text{geom}}$.
}
\label{fig:GNoMEs_EMD_PDA10_ICSD_MP}
\end{figure}

%B.3---------------------
\subsection{Methods of comparing GNoMEs with ICSD and MP}
\label{subsection:methods}

The comparison of GNoMEs proceeds in the following steps. The query set is the union of GNoME1 and GNoME2 totalling $637{,}579$ structures (retained entries were compared once and reported in both releases). The reference sets were the ICSD and the Materials Project. Counts are of the union of the two releases against each reference set. Justifications for each step and distance are in Table~\ref{tab:comparison_distances}.

\begin{enumerate}

\item \emph{Candidate lists by $\ADA$}. For each entry and each reference set, three lists ranked by RMS on $\ADA[100]$: the five nearest references overall, the five nearest sharing its element set, and the twenty nearest sharing its element set and having close enough composition. The third list is the only route by which an ordered supercell of a disordered crystal reaches the comparison, as its $\ADA$ is not similar to any one configuration.

\item \emph{$\PDA$ comparison of the candidates}. Each structure is compared by its $\PDA[100]$. Every candidate pair within an $\ADA$ gate of $0.05$, widened to $0.2$ for a same-composition candidate, is compared for $d_{\text{geom}}$ and $d_{\text{chem}}$. Pairs with the same composition where the reference has partly vacant sites has $d_{\text{fw}}$ computed as well.

\item \emph{Reading the EMD plan site by site}. For each of the $3{,}280$ pairs whose reference is disordered, shares the entry's element set and composition, and has $d_{\text{chem}}$ within $0.2$, the chemically constrained plan is re-read: for every mixed site of the reference that receives flow, which elements the entry places there and in what shares, and the distance between the closest two rows of different elements sitting on it. Those are the \emph{witness} distances of the following subsection. 

\item \emph{Per-entry distances and categories}. A pair is a chemical candidate when the entry shares the reference's element set and composition. Such a pair scores $d_{\text{chem}}$ and, where the reference has partly vacant sites and the surviving framework (after deleting vacancies) retains at least half of its atoms, $\min(d_{\text{fw}}, d_{\text{chem}})$. Each entry then takes the smallest such distance over all of its pairs, separately for ordered and for disordered references, and is assigned the first category that comes within $\tau$: \emph{ordered}, then \emph{disordered}, then \emph{geometry only} at $d_{\text{geom}}$, otherwise none. All disordered matches are combined into one \emph{disordered} category for the results section.
\end{enumerate}

\begin{table}[htbp]
  \centering
  \caption{Comparison screening steps. Each establishes something the one above it cannot, and the third column gives the case that forces the next. Distances are RMS $\PDA[100]$ EMD in \AA ngstr\"oms. The last 3 are combined in results into ``disordered''.}
  \label{tab:comparison_distances}
  \setlength{\tabcolsep}{4pt}
  \small
  \begin{tabular}{@{}p{0.155\textwidth}p{0.30\textwidth}p{0.47\textwidth}@{}}
    Screen & Establishes & Why it is not sufficient alone \tabularnewline
    \hline
    \raggedright $d_{\text{geom}}$ \newline \emph{geometry only}
      & \raggedright The two point sets agree as geometry.
      & \raggedright It says nothing about chemistry. \tabularnewline[3pt]
    \raggedright $d_{\text{chem}}$, ordered reference \newline \emph{ordered}
      & \raggedright A transport plan exists that moves mass only between the same elements.
      & \raggedright Sufficient here, and sharing an element set is not: CoFe$_2$ (95369b14fd) aligns with Co$_5$Fe$_{11}$ (ICSD 155842) to $0.0053$, but $56\%$ of the mass would have to cross Co$\leftrightarrow$Fe, so $d_{\text{chem}} = \infty$. \tabularnewline[3pt]
    \raggedright $d_{\text{chem}}$, disordered reference
      & \raggedright The same, with a mixed reference site accepting flow from any element that occupies it.
      & \raggedright Necessary but weak because a mixed site accepts either element. B$_2$FeNb$_2$Ru$_5$ (dbf7583a57) is distance $0.0059$ from B$_2$Fe$_{0.83}$Nb$_{2.17}$Ru$_5$ (ICSD 195034), sending only Fe to the (Fe$_{0.83}$Nb$_{0.17}$) site. \tabularnewline[3pt]
    \raggedright witness, or every mixed site resolved \newline \emph{ordering}
      & \raggedright Two rows of the entry carrying different elements reach one mixed site, so exchanging their labels leaves the $\PDA$ unchanged; the pair has distance $\max(d_{\text{chem}}, \text{witness})$. Failing that, every mixed site takes a single element.
      & \raggedright Needed because the shares can be right and the geometry wrong: B$_6$NbNi$_{20}$U$_2$ (f82eb9dc20) puts Nb$_{0.5}$U$_{0.5}$ on the (Nb$_{0.534}$U$_{0.466}$) site of ICSD 011919, yet those two rows lie $0.414$ apart. The across-sites form is weaker: with the two elements on inequivalent positions there is no label swap to appeal to, only the composition gate. \tabularnewline[3pt]
    \raggedright $d_{\text{fw}}$ \newline \emph{framework}
      & \raggedright For a reference with partially occupied sites: the vacancy atom types are deleted from both crystals and the surviving ordered sublattice compared.
      & \raggedright It is not a relaxation of $d_{\text{chem}}$ but the only finite answer for these, since deleting an atom perturbs every neighbouring row. Guarded by requiring the framework to keep half the reference's atoms: stripping O and Gd from Gd$_6$O$_{12}$U (ICSD 021945) leaves one U atom of 19. \tabularnewline
    \hline
  \end{tabular}
\end{table}

%B.4--------------------------
\subsection{Details of the most striking examples in GNoMEs 1\&2}
\label{sub:GNoMEstriking}

This section provides more details about invariants that detect outliers in GNoMEs, most of which seem to be physically unrealistic.
The two contact bounds $\dmin$ and $\dmax$ are not equally decisive. 
GNoME2's largest $\dmax$ is $5.29\angstrom$ and the next largest is 4.63\AA{}, so the 50 GNoME1 entries above it, running up to 8.92\AA for entry eed3027c1e of composition \ce{KGd(Cu2S)2}, are well-separated from GNoME2. 
The lower contact bound is not like this. 
GNoME2's smallest $\dmin$ is 0.7482\AA, and apart from the two Na$_4$K entries with atoms on the same positions in \cite[Fig.~15]{anosova2026recognition}, the whole ``atoms too close'' set is 21 hydrides sitting between 0.7460 and 0.7481\AA: H$_6$Fe, H$_8$Cr, H$_8$Co, H$_8$Cu, H$_8$Be, H$_8$Ga etc, all of them the H--H separation of a hydrogen molecule (0.74\AA). 
Both GNoME 1\&2 have many such contacts. 
GNoME1's just fall a few thousandths of an $\angstrom$ on the wrong side of GNoME2's own extremum. 
All 23 were removed regardless.
\medskip

\begin{table}[h!]
  \centering
  \caption{The ten sparsest and the five most overlapped GNoME1 entries. All were removed from GNoME2. 
  Here $\rho$ is the point density, and $\dmin$ and $\dmax$ are the smallest and largest distances between any atoms. The median GNoME1 entry has $\rho = 52.2$ nm$^{-3}$. The two Na$_4$K entries are the same structure twice.}
  \label{tab:implausible-examples}
  \setlength{\tabcolsep}{3pt}
  \begin{tabular}{llrrrr}
    entry & formula & $\rho$ (nm$^{-3}$) & $\dmin$ (\AA) & $\dmax$ (\AA) & density (g/cm$^3$) \\
    \hline
    \multicolumn{6}{l}{\emph{sparsest}} \\
    88e5e5c7a3 & S$_7$Gd$_2$Tb$_3$      & 1.29 & 4.42 & 8.30 & 0.18 \\
    c247c290e6 & Cu$_3$Sn$_2$I$_9$U$_2$ & 1.73 & 6.90 & 7.58 & 0.37 \\
    37f9f79871 & P$_3$FeCo$_2$U$_2$     & 1.81 & 7.87 & 8.48 & 0.28 \\
    ea4f390476 & Ge$_2$GdPt$_2$Th       & 1.94 & 6.62 & 7.73 & 0.50 \\
    ce080d08df & OCu$_2$La$_9$Pb$_6$    & 2.31 & 4.34 & 8.15 & 0.56 \\
    f4614ac0c2 & Ni$_3$In$_3$CeGd$_2$   & 2.32 & 7.04 & 8.25 & 0.42 \\
    784a1426b2 & Mg$_2$Ga$_3$GdTb$_3$   & 2.38 & 7.09 & 8.11 & 0.39 \\
    bd207edbb8 & S$_9$Gd$_2$Tb$_2$      & 2.40 & 3.75 & 7.39 & 0.28 \\
    ca5372206f & SnSmGdAu$_2$           & 2.49 & 6.61 & 7.93 & 0.68 \\
    cb590a0692 & Ga$_4$Nd$_4$GdPt$_3$   & 2.66 & 6.84 & 7.94 & 0.59 \\
    \hline
    \multicolumn{6}{l}{\emph{most overlapped}} \\
    0e2d8f26d6 & Na$_4$K                & 21.9  & 0.000 & 4.20 & 0.95 \\
    cdc06a1a2a & Na$_4$K                & 21.9  & 0.000 & 4.20 & 0.95 \\
    98cb8c7d70 & H$_8$Li$_2$CsIr        & 71.8  & 0.746 & 2.96 & 3.45 \\
    a1590781f0 & H$_6$Fe                & 116.4 & 0.746 & 1.74 & 1.71 \\
    ba6e08bb47 & H$_8$Cr                & 116.1 & 0.746 & 1.80 & 1.29 \\
  \end{tabular}
\end{table}

\begin{figure}
    \centering
    \includegraphics[width=1\linewidth]{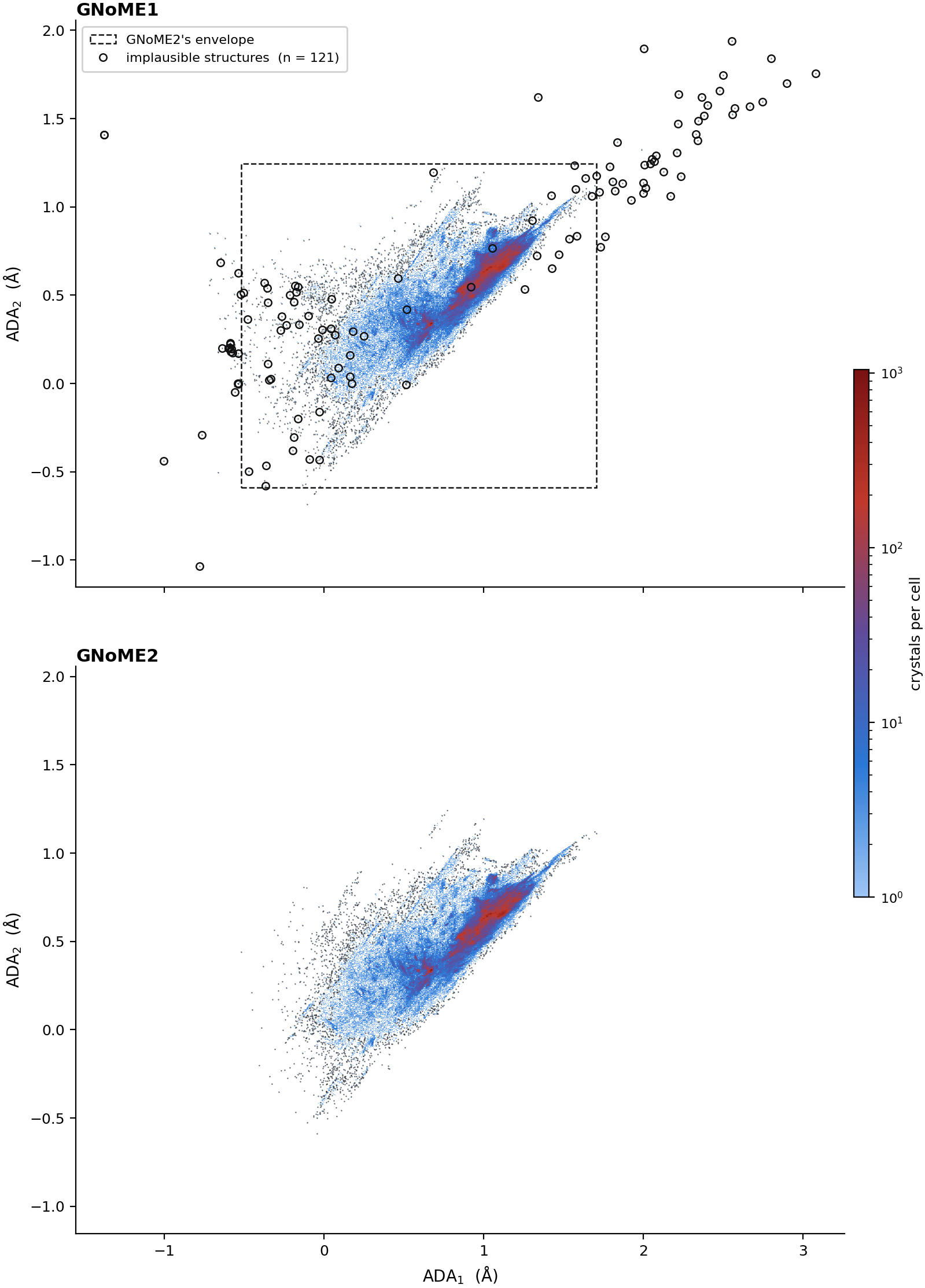}
    \caption{GNoMEs 1\&2 are plotted on the axes $\ADA_1$ vs $\ADA_2$, with the same scale. The most unusual entries that were removed are small circles sitting far from the main cloud in GNoME1.}
    \label{fig:placeholder}
\end{figure}

\begin{table}[htbp]
  \centering
  \caption{GNoME1 entries falling outside the range GNoME2 occupies. Each bound is GNoME2's own extremum, so no threshold is chosen by hand. The categories overlap, 41 of the 121 entries violate two bounds and 13 violate three, so the counts sum to more than the union. The background removal rate is 21.7\% (83,525 of 384,938).}
  \label{tab:implausible}
  \begin{tabular}{llrrr}
%    & & entries in & removed \\
    violated bound & criterion & entries in GNoME\,1 & removed (\%) \\
    \hline
    too sparse          & $\rho < 12.61$ nm$^{-3}$    & 94  & 100.0 \\
    atoms too close    & $\dmin < 0.7482$ \AA        & 23  & 100.0 \\
    atoms too far  & $\dmax > 5.29$ \AA          & 50  & 100.0 \\
    density too low    & $< 0.49$ g/cm$^3$           & 21  & 100.0 \\
    \hline
    any of the above   &                             & 121 & 100.0 \\
  \end{tabular}
\end{table}

\begin{figure}
    \centering
    \includegraphics[width=\linewidth]{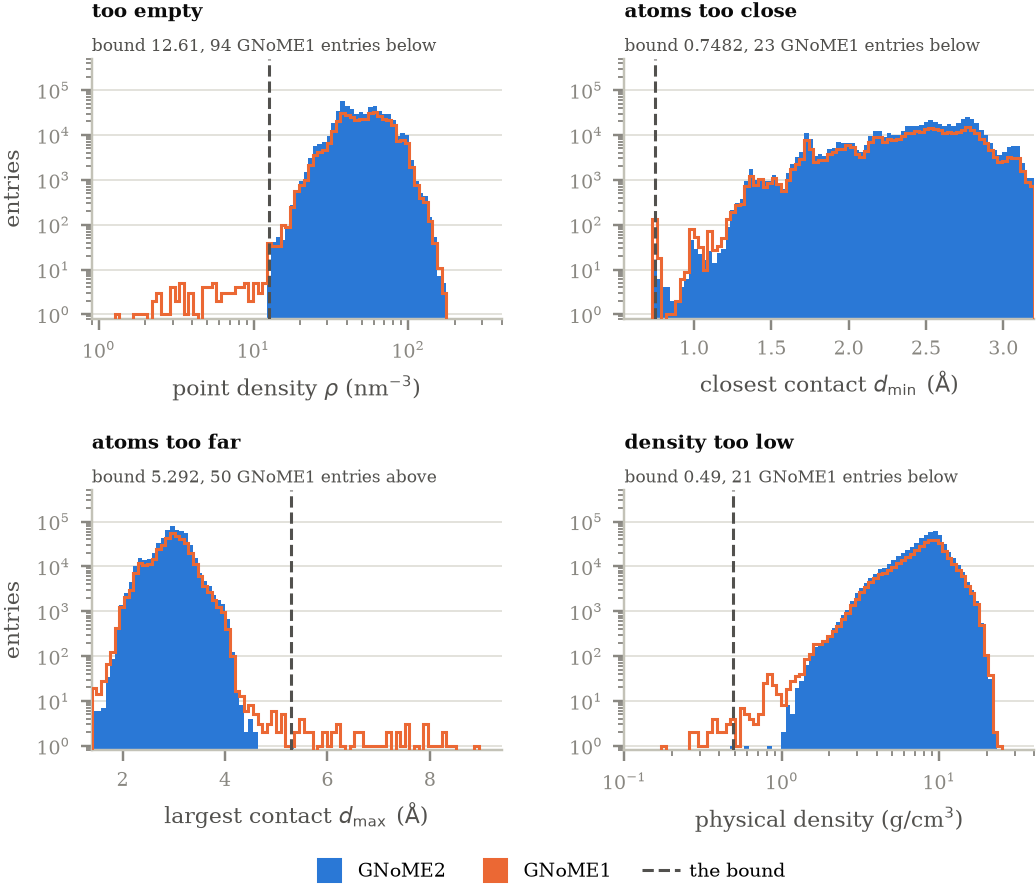}
    \caption{The distribution of entries in GNoME1 and GNoME2 by the four values used to determine implausible structures in Table~\ref{tab:implausible}.}
    \label{fig:implausible}
\end{figure}

All 121 entries are listed in the supplementary file \texttt{implausible\_structures.csv} with the bound each one violates.
Of the 121 implausible removed structures, 44 went two days after the initial release, and the remaining 77 nine months later. 
The 12/1/23 step is striking in that it withdrew only 67 entries, and 44 of them are entries these bounds independently flag.  Both Na$_4$K entries, the only structures with superimposed atoms, went in the first wave, while 21 of the 23 hydrides survived. 
The emptiness flags divided more evenly in the sense that it did not specifically target empty cells, and the entries it did remove were the more extreme ones: their median point density is 5.96 nm$^{-3}$ against 12.09 nm$^{-3}$ for those that lasted until 8/21/24.

\begin{table}[!h]
\centering
\begin{tabular}{lllll}
GNoME ID & ICSD ID & GNoME composition & ICSD composition & $\EMD$ on $\PDA[100]$ \\ \hline
cdc06a1a2a & 138567  & KNa4        & I3V       & 0.649084 \\
0e2d8f26d6 & 138567  & KNa4        & I3V       & 0.649084 \\
eed3027c1e & 281367  & Cu4GdKS2    & Y         & 0.623388 \\
3e5dc620b0 & 27030   & GdIn2Pt4Rh  & S         & 0.584775 \\
ce080d08df & 20564   & Cu2La9OPb6  & CoK       & 0.565175 \\
%bf4c20274a & 248434  & CeNiPtSb2   & F2Sr      & 0.543186 \\
%37f9f79871 & 29124   & Co2FeP3U2   & P         & 0.535419
%b9fd231415 & 60570   & CsEuNiS     & CdS       & 0.532303 \\
%f4614ac0c2 & 687551  & CeGd2In3Ni3 & AlFeNi2   & 0.524135 \\
%348033e3a9 & 110271  & Cl7Nb3Pu    & C3H14Cl5N & 0.455544 \\
\end{tabular}
\caption{Top 5 predicted crystals $S$ in GNoME1 that maximise the Local Novelty Distance $\LND\langle \EMD,\PDA[100], \text{ICSD}\rangle(S)$ to their nearest neighbour in the ICSD.
The first two predictions cdc06a1a2a and 0e2d8f26d6 include two pairs of repeated atoms, see \cite[appendix~A, Fig~15]{anosova2026recognition}.
The 3rd prediction eed3027c1e has the largest $\dmax=8.92\angstrom$ in Fig.~\ref{fig:GNoME1_largest_dmax_dmin}~(left).
The next two predictions in Fig.~\ref{fig:GNoME_LND_EMD_PDA100_ICSD} also have no close atomic contacts.
}
\label{tab:GNoME_LND_EMD_PDA100_ICSD}
\end{table}

\begin{figure}[h!]
\centering
\includegraphics[height=40mm]{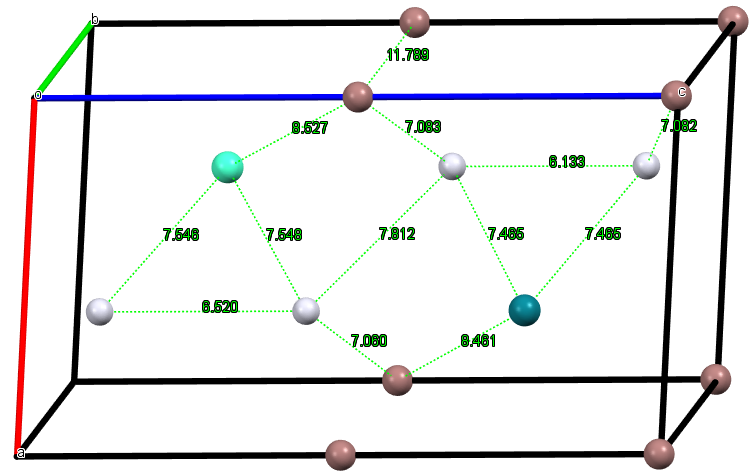}
%\hspace*{1mm}
\includegraphics[height=40mm]{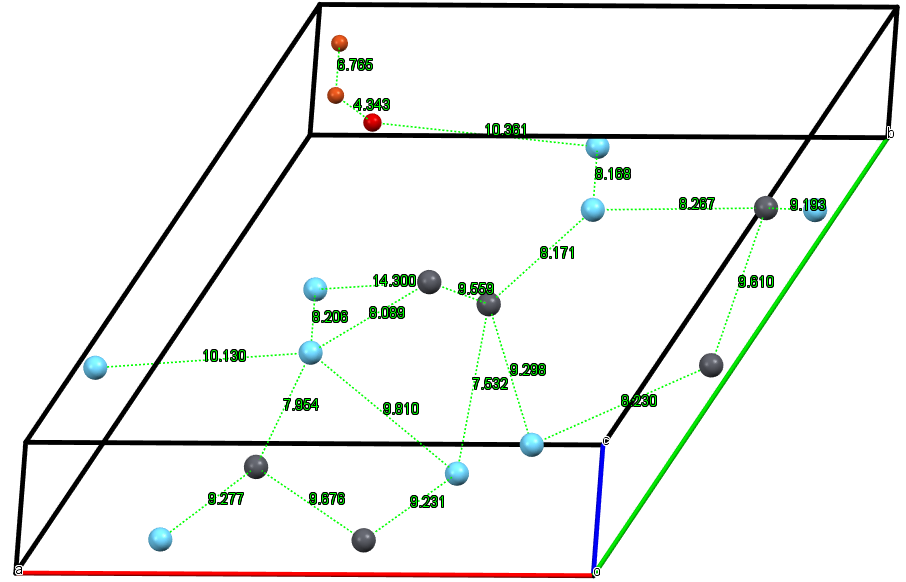}
\caption{
Unrealistic predictions in GNoME1 with large values of $\LND\langle \EMD,\PDA[100], \text{ICSD} \rangle$ in Table~\ref{tab:GNoME_LND_EMD_PDA100_ICSD} and
without close atomic contacts.
\textbf{Left}: entry 3e5dc620b0. 
\textbf{Right}: entry ce080d08df.
}
\label{fig:GNoME_LND_EMD_PDA100_ICSD}
\end{figure}

%C======================
\section{More details of software for crystal disguising}
\label{sec:crystal_disguise}

The second author developed a web application in Python called ``Disguise a Crystal''.
The underlying white-box algorithm can generate 2.2 million new-looking crystals in about 24 hours on a modest laptop with the power of school-level geometry rather than AI.
With the press of a single button, anyone can disguise an existing crystal so that its Crystallographic Information File (CIF) looks completely different. 
After ``disguising" the CIF, most if not all the numbers in the file change and it is difficult to manually spot that the crystal it represents is either an exact or near-duplicate. 
\medskip

Filters that do not use Lipschitz continuous metrics cannot recognise the disguised CIF as a near-duplicate reliably.
For example, Mercury's Crystal Packing Similarity tool in the 2024.2 CSD Release often crashes when attempting to compare the original file and its disguise, presumably due to the exponentially many bijections between chemically identical atoms.
This problem for continuously comparing finite atomic configurations under rigid motion has been resolved in polynomial time in \cite{kurlin2027complete}.  
\medskip

The ``Disguise a Crystal'' program motivates the necessity for rigorous and computationally fast methods for crystal comparison. 
It is important to analyse new creations to see which are the same and which are different. 
It is not enough to generate millions of files using AI and claim that they are crystals that are new or in fact real.
\medskip

The simplest trick is to randomly permute (lines of) atoms in a given CIF, which already creates exponentially many different representations of crystals with indistinguishable atoms.
Any attempt to order atoms by coordinates breaks down continuity under small perturbations.
For example, a lexicographic order of triples $(x,y,z)$ requires that $(0,0.3,0.6)<(0,0.3,0.9)$, but a slight change in the $y$ coordinate gives the opposite order: $(0,0.29,0.9)<(0,0.3,0.6)$.
\medskip
  
Even if all given atoms are unique, an initial cell can be transformed by a random $3\times 3$ matrix that has integer coefficients and determinant 1.  
The resulting skewed cell can be arbitrarily extended by any integer factors along the new basis vectors.
This extension also enlarges a motif of atoms, which complicates any further processing without changing the underlying structure (a class under rigid motion). 
In theory, the two transformations above can be neutralised by reducing the cell to (say) Niggli's form.
In practice, any slight perturbation of at least one atom in the new cell (and all its repeated copies by lattice translations) makes the resulting extended cell primitive, and any reduction of the new cell has the same arbitrarily increased volume. 
\medskip

Even worse, if any chemical elements are replaced without changing their geometric positions or by only tiny perturbations, the new CIF can look completely different from the original.
If we extend a given unit cell by a factor of 1 or 2 along each of the three basis vectors ($2^3-1=7$ non-trivial extensions) and then independently shift 10 times one of (say) 10 atoms in the extended cells, we can get $7\times 10^{10}$ (70 billion) numerically different CIFs, even without extra disguises such as permuting atomic rows and changing chemical elements.
\medskip

Replacing a chemical element with its neighbour in the same group of Mendeleev's table might only slightly affect structure factors if such a crystal can really be synthesised. 
The resulting crystal can look novel and realistic despite being a disguised near-duplicate of a previously known material. 

%C.1-----------------
\subsection{How to Disguise your own Crystal}

You can run the program on your own computer by downloading the files from the GitHub Repository \cite{kurlina2026disguise}. The `readMe' file guides you on how to download and run the program for Windows and Mac.
\medskip

\begin{figure}[h!]
    \centering
    \includegraphics[width=\textwidth]{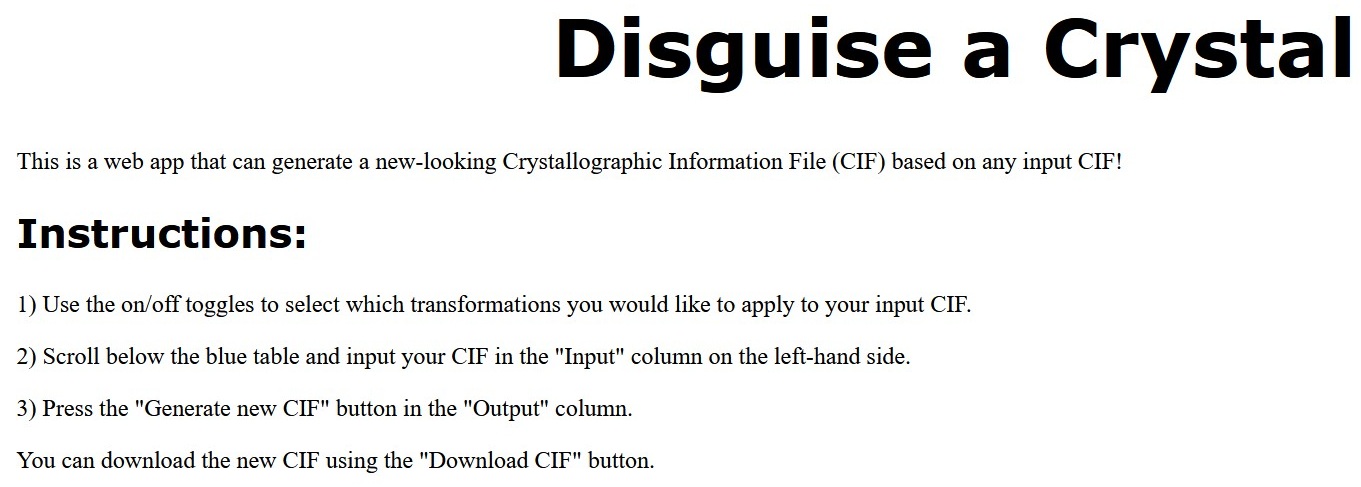}
    \includegraphics[width=\textwidth]{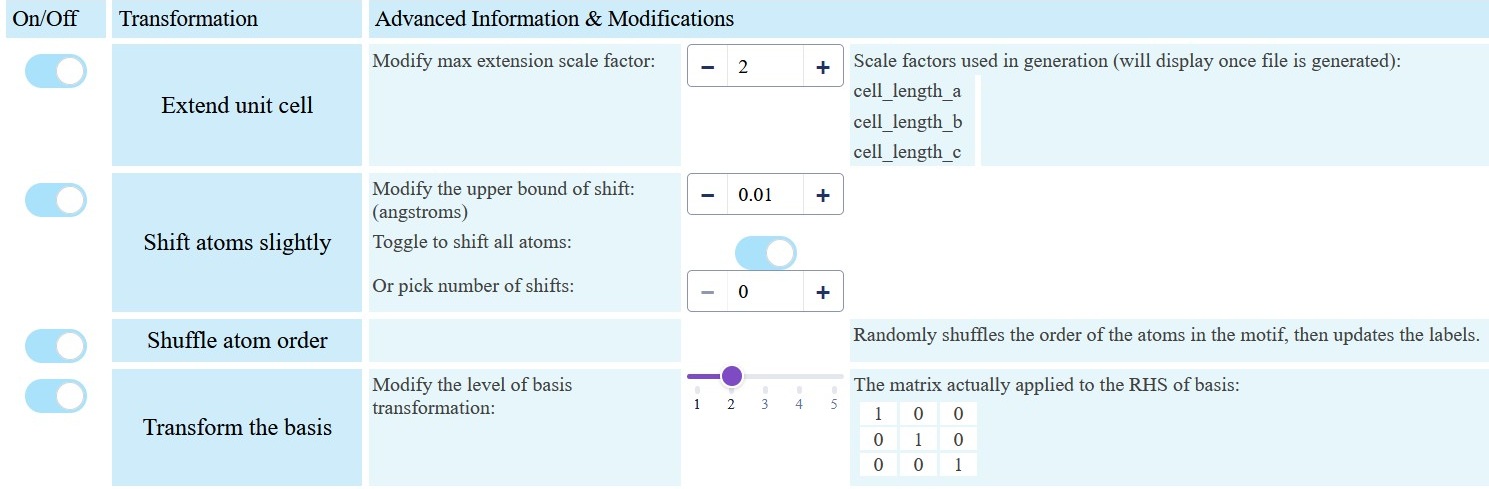}
    \includegraphics[height=68mm]{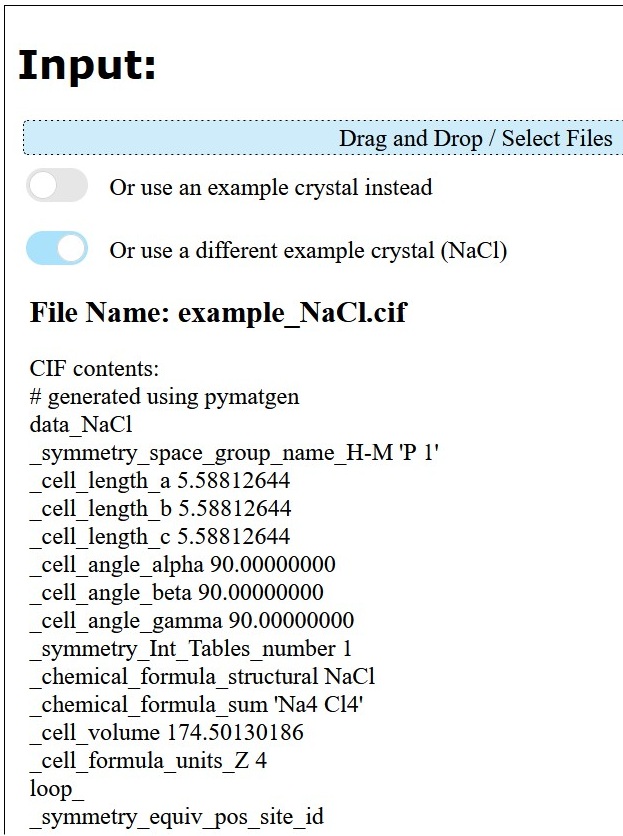}
        \includegraphics[height=68mm]{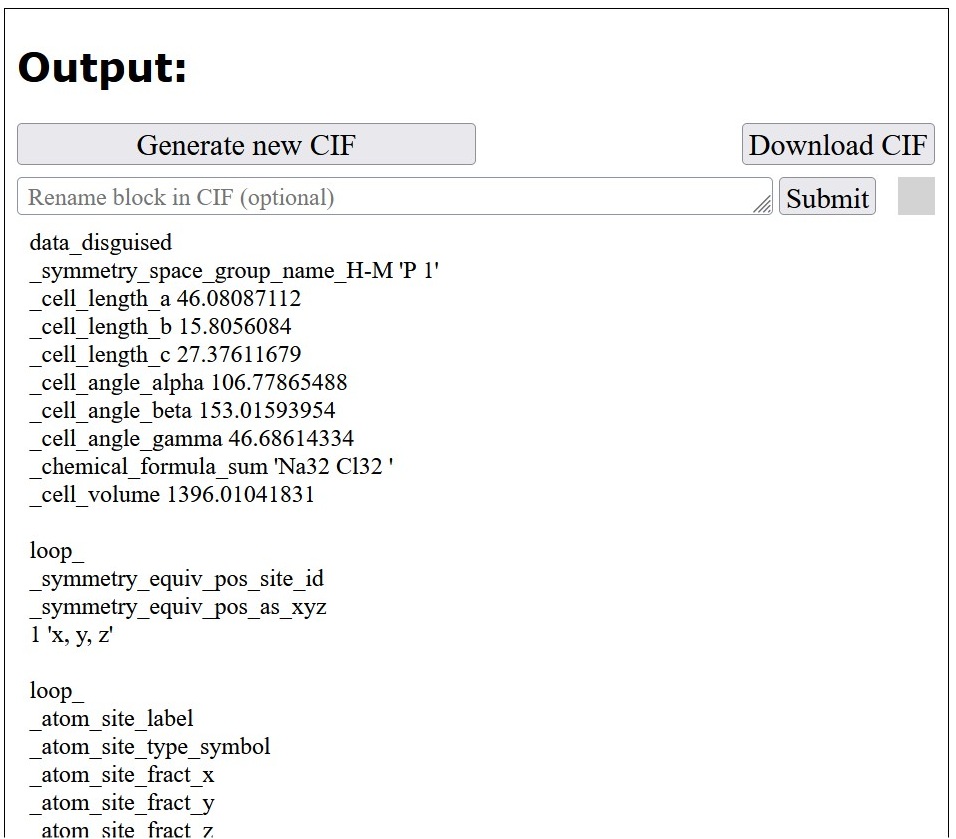}
    \caption{Screenshots of the user interface and example inputs.}
\end{figure}

First, select which settings you would like to run the program with and which transformations you would like to apply to disguise the crystal using the interactive buttons and input fields in the blue table in the User Interface.
\medskip

Next, upload your own CIF to the Input column below the table or use the examples provided 
and once you press the ``Generate CIF'' button, the output CIF will be displayed and 
be available to download.

%C.2-----------------
\subsection{Algorithm Details}

The algorithm inputs a correctly formatted CIF (Crystallographic Information File) \cite{cifFormat1991} 
with unit cell information and coordinate data listed in the first block. 
The crystal in the CIF may be in any space-group but should have no disorder which will be ignored by the program. 
The algorithm consists of 4 independent transformations:

\begin{enumerate}
\item \emph{Extension of the Unit Cell.} 
The program randomly picks 3 scale factors up to the upper bound (2 by default) and extends each unit cell length correspondingly. 
All atoms in the original motif are duplicated accordingly to fill the new, larger unit cell.

\item \emph{Tiny perturbations of atom coordinates.} Atoms are converted to Euclidean Coordinates, and are shifted by a vector which is sampled uniformly from a 3D ball using the algorithm from  \cite[section~3.29]{fishman2013monte}. 
The default maximum shift is $0.01$ Angstroms but this can be changed in the ``Advanced Information \& Modifications'' section.

\item \emph{Change of basis.} The unit cell information is orthogonalised i.e. converted into $3$ basis vectors, $b_1,b_2,b_3$ in the Euclidean Space, $\mathbb{R}^3$, that define the unit cell. 
The basis vectors are then transformed by applying a generated $3$ by $3$ matrix on the right. 
The generated matrix has determinant $1$ and is formed by first randomly generating $a,b,c,d,e,f \in \mathbb{Z} $ in the range $(-k,k)$ where $k$ is the level of basis transformation that can be modified in settings, and then forming the matrix:
\begin{align*} 
    \begin{pmatrix}1 & a & b\\0 & 1 & c\\0 & 0 & 1\end{pmatrix}
    \times \begin{pmatrix}1 & 0 & 0\\d & 1 & 0\\e & f & 1\end{pmatrix} =
    \begin{pmatrix}1+a d+be & a+bf & b\\d +ce & 1+ c f& c\\e & f & 1 \end{pmatrix} 
    \end{align*} 
    The new basis is:
    \begin{align*} 
    \begin{pmatrix}\vert & \vert& \vert\\b_1&b_2&b_3\\\vert &\vert & \vert\end{pmatrix} 
    \begin{pmatrix}1+a d+be & a+bf & b\\d +ce & 1+ c f& c\\e & f & 1 \end{pmatrix} 
    \end{align*} 
    The new fractional coordinates are found by applying the inverse of the generated matrix on the left of the original fractional coordinate. If the original fractional coordinate $(x,y,z)$, the new coordinate vector is:
    \begin{align*} 
    \begin{pmatrix}1+a d+be & a+bf & b\\d +ce & 1+ c f& c\\e & f & 1 \end{pmatrix} ^{-1}
    \begin{pmatrix}x\\y\\z\end{pmatrix}
\end{align*} 
This way, the euclidean coordinates remain the same while the unit cell lengths and angles, and fractional coordinates are different.

\item \emph{Shuffling of the atomic order.} 
The atoms in the unit cell are arbitrarily ordered and hence can be randomly shuffled.
Atomic labels are updated so that the labels are still listed chronologically in the final file.
\end{enumerate}	

\medskip
The output of the program is a correctly formatted CIF that can be processed by current crystal software that handle CIFs.

\medskip

If the default transformations were applied, most if not all the numbers in the output file will be different from the numbers in the input file. However, the crystal structure represented will be either the same geometrically or very slightly different.
\medskip

If no perturbations were applied, then the Earth Mover's Distance (EMD) on the Pointwise Distance Distributions (PDDs) of the input and output files will be $0$, or less than $10^{-4}$ Angstroms, which is due to floating point calculations and is insignificant \cite{widdowson2026pointwise}. 
If perturbations up to $0.01$ Angstroms were applied to all atoms, the maximum EMD is $0.02$ Angstroms.
\medskip

All the elements are kept the same, so if the input CIF represents a real crystal, the output CIF is guaranteed to represent a real crystal too. 
\medskip

For full transparency, comments on lines starting with \# are added to the bottom of the input CIF that list the settings of the program when the file was generated and also the transformation matrix and the extension that took place.	

\end{document}